\documentclass{elsart}

\usepackage{amsmath}
\usepackage{graphicx}

\begin{document}

\begin{frontmatter}



\title{Fluctuations in fluids in thermal nonequilibrium states
below the convective Rayleigh-B\'{e}nard instability}
\author[label1]{Jos\'{e} M. Ortiz de Z\'{a}rate}
\address{Departamento de F\'{\i}sica Aplicada I, Facultad de Ciencias F\'{\i}sicas, Universidad Complutense, E28040 Madrid, Spain}
\author{Jan V. Sengers}
\address{Institute for Physical Science and Technology and Department of Chemical
Engineering, University of Maryland, College Park, MD 20742, USA}
\thanks[label1]{Corresponding author}


\begin{abstract}
Starting from the linearized fluctuating Boussinesq equations we derive
an expression for the structure factor of fluids in stationary
convection-free thermal nonequilibrium states, taking into account
both gravity and finite-size effects. It is demonstrated how the
combined effects of gravity and finite size causes the structure factor
to go through a maximum value as a function of the wave number $q$.
The appearance of this maximum is associated with a crossover from
a $q^{-4}$ dependence for larger $q$ to a $q^2$ dependence for very
small $q$. The relevance of this theoretical result for the interpretation
of light scattering and shadowgraph experiments is elucidated. The
relationship with studies on various aspects of the problem by other
investigators is discussed. The paper thus provides a unified treatment
for dealing with fluctuations in fluid layers subjected to a stationary
temperature gradient regardless of the sign of the Rayleigh number $R$,
provided that $R$ is smaller than the critical value $R_\mathrm{c}$
associated with the appearance of Rayleigh-B\'{e}nard convection.
\end{abstract}

\end{frontmatter}

\maketitle

\section{Introduction}
\label{sec:level1}
Questions concerning the nature of thermal fluctuations in fluids in
thermal nonequilibrium states have been the subject of many studies
during the past two decades. Specifically, in this paper we consider
thermal fluctuations in a horizontal layer of a one-component fluid that
is heated either from above or from below in such a way that the fluid
is still in a stable convection-free state. A number of different and
independent approaches for dealing with various aspects of this problem
have appeared in the literature.

Most of the earlier studies focused on the nonequilibrium fluctuations
close to the convective Rayleigh-B\'{e}nard instability because of their
possible influence on the pattern-selection process when convection
appears~\cite{ZaitsevShliomis,Graham,GrahamPleiner,SwiftHohenberg}.
The effects of both linear and nonlinear terms in the hydrodynamic
fluctuations close to the convection threshold have been evaluated.
This line of research turned out to be very important, mainly because
it lead to the development of convection model equations for dealing
with spatiotemporal-pattern formation above the threshold, as
reviewed by Cross and Hohenberg~\cite{CrossHohenberg}.

A second and independent line of research developed when investigators
became interested in the nature of the fluctuations in nonequilibrium
steady states far away from any hydrodynamic instability. The first
important theoretical work along this line was done by Kirkpatrick et al.,
who calculated the correct expression for the structure factor of a fluid
subjected to a stationary temperature gradient~\cite{KirkpatrickEtAl}.
Althought their result was originally obtained on the basis of the same
mode coupling theory that in equilibrium leads to long-time tail
contributions to the correlation functions for the transport coefficients,
it turned out that the same result could also be obtained from the
simpler scheme of Landau's fluctuating hydrodynamics~\cite{RonisProcaccia,SchmitzCohen1,LawSengers}.
This line of research received an important momentum from the experimental
side when the research group of Sengers and coworkers verified from
light-scattering experiments the main conclusion of the theoretical
result, namely that the nonequilibrium contribution to the structure
factor of a liquid is proportional to the square of the temperature
gradient $\nabla T$ and inversely proportional to the fourth power
of the wave number $q$ of the fluctuations~\cite{LawGammonSengers,LawEtAl,SegreEtAl1}.
Theory and experiments have subsequently been extended to also deal with
nonequilibrium fluctuations in liquid mixtures and polymer solutions,
as reviewed elsewhere~\cite{LectureNotes}.

A third line of independent research has focused on the real-space
behavior of the nonequilibrium correlations resulting from the
heat-diffusion equation~\cite{GarciaEtAl1,RubiEtAl,LiuOppenheim,GarciaEtAl2}.
A study of the correlations resulting from the heat-diffusion
equation is equivalent to a study of the nonequilibrium
correlations in a liquid in the direction co-incident with
the temperature gradient. This line of research has shown how
the nonequilibrium correlation function of the fluctuating variables
encompasses the entire system not involving any intrinsic lenght
scales. The real-space analysis has demonstrated the long-range nature
of the nonequilibrium fluctuations in the direction of the
temperature gradient, which has been further confirmed
by numerical integration of the equations~\cite{GarciaEtAl1},
by computer simulations~\cite{MalekMansourEtAl2} and by a
lattice-gas automaton approach~\cite{SuarezEtAl}.

It is our present goal to further extend the second aforementioned
research so as to describe nonequilibrium fluctuations far away
from any convective instability, nonequilibrium fluctuations close
to the convective instability and nonequilibrium fluctuations produced
by the heat-diffusion equation from one unified point of view. The
original work of Kirkpatrick et al. and of others~\cite{KirkpatrickEtAl,RonisProcaccia,LawSengers}
yielded an expression for the structure factor of a fluid in thermally
nonequilibrium states without considering any gravity or boundary
effects. Segr\`{e} at al. extended the theory to include gravity
effects, first in a one-component liquid~\cite{SegreSchmitzSengers}
and then in a liquid mixture~\cite{SegreSengers}. They predicted
that gravity will cause the singular $q^{-4}$ dependence of the structure
factor to saturate at small values of $q$. This prediction was confirmed
experimentally by Vailati and Giglio from light-scattering experiments
at ultra-small scattering angles~\cite{VailatiGiglio1,VailatiGiglio2}.
More recently, Ortiz de Z\'{a}rate et al. evaluated the nonequilibrium
structure factor including finite-size effects but neglecting
gravity~\cite{Physica,EPJ}. They concluded that deviations from the
$q^{-4}$ dependence due to finite-size effects will be just as important
as deviations due to gravity. Hence, for a quantitative interpretation
of ultra-low-angle light scattering or shadowgraph experiments it is
important to account for both gravity and finite-size effects.

In this paper we extend the previous work so as to include gravity and
finite-size effects simultaneously. Starting from the linearized fluctuating
Boussinesq equations, we shall derive a complete expression for the
structure factor of the nonequilibrium fluid making no other approximations
than the ones incorporated in the Boussinesq equations. We shall then show
that the resulting expression is capable of describing nonequilibrium
fluctuations both far away and close to the convective instability.
Close to the convective instability we shall recover the linear
Swift-Hohenberg correlation function as an approximation to our more
complete result. From the study of the behavior of the nonequilibrium
correlation function in real space, we shall recover with appropriate
approximations, expressions and plots previously obtained by other
investigators for the case of the heat-diffusion equation. Thus it will be
shown how these various conclusions can be reached from one single theoretical
result.

Attempts to arrive at a unified description of nonequilibrium fluctuations
both far away and close to the convective instability has been made by
Kirkpatrick and Cohen~\cite{KirkpatrickCohen} and by Schmitz and
Cohen~\cite{SchmitzCohen2}, but the consequences of their theoretical
results for the interpretation of actual experiments are not clear.
In contrast, we shall be able to make a connection with available
experimental data for the thermal fluctuations near the convective
Rayleigh-B\'{e}nard instability~\cite{WuEtAl}.

There has been an extensive debate whether fluctuating hydrodynamics
can properly account for the thermal fluctuations close to the onset
of convection~\cite{AhlersEtAl,MeyerEtAl1,VanBeijerenCohen,AhlersEtAl2,MeyerEtAl2},
but the most recent answer to this question appears to be
positive~\cite{WuEtAl,RehbergEtAl}.
The present paper will
provide additional evidence for the validity of fluctuating hydrodynamics
to describe nonequilibrium fluctuations close to the onset of the
classical Rayleigh-B\'{e}nard instability in simple fluids.

We shall proceed as follows. We start from the linear fluctuating
Boussinesq equations which will be written in a form most suitable
for our analysis in Section~\ref{sec:0}. In Section~\ref{sec:1} we review
the derivation of the well-known expression for the structure factor
of a fluid subjected to a stationary temperature gradient without taking
into account the presence of boundaries, but including the effects of
gravity. The relationship of the resulting expression for the effect of
gravity with that obtained previously by Segr\`{e} et al.~\cite{SegreSchmitzSengers}
will be elucidated. In Section~\ref{sec:2} we then consider the modifications
to the nonequilibrium structure factor due to the finite height of
the fluid layer. In Section~\ref{sec:LS} we present a detailed analysis
of how the combined effects of gravity and finite-size will appear in
low-angle light scattering and in shadowgraph experiments.
Sections~\ref{sec:2} and~\ref{sec:LS} contain the most important
new features of the present work. In Section~\ref{sec:INS} we introduce
an approximation valid close to the convective instability and show how
one can recover the linear Swift-Hohenberg model for the fluctuations near the
convective threshold. In Section~\ref{sec:RS} we discuss the spatial behavior
of the nonequilibrium structure factor and we shall make contact with the
results obtained by previous investigators for the long-range nature
of the nonequilibrium fluctuations in the direction of the
temperature gradient. Our results are summarized in
Section~\ref{sec:CON}.

\section{Linearized fluctuating Boussinesq equations}
\label{sec:0}
We consider a fluid layer between two horizontal plates separated by a
distance $L$. The fluid layer is subjected to a temperature gradient in
the vertical direction by maintaining the plates at two different
temperatures. The size of the system in the two horizontal
X- and Y-directions is much larger than the size $L$ in the
vertical Z-direction.

To determine the structure factor of the fluid we consider small
fluctuations around the conductive solution. These small fluctuations
shall be described by the linearized Boussinesq equations
supplemented with random noise terms, as first considered by
Zaitsev and Shliomis and by Swift and Hohenberg for studying the influence
of thermal noise close to the convective instability~\cite{ZaitsevShliomis,SwiftHohenberg}.
Use of the Boussinesq approximation to the full hydrodynamics equations
implies that we neglect the sound modes and consider only density
fluctuations caused by temperature fluctuations. The temperature gradient
is applied along the Z-direction and is written in the form:
\begin{equation}
\boldsymbol{\nabla}T_0=\nabla T_0~\hat{z},
\end{equation}
where $T_0=\langle T \rangle$ is the average local value of the temperature
$T$ and where $\nabla T_0 = dT_0/dz$. Hence, $\nabla T_0$ is positive when
the fluid layer is heated from above and negative when heated from below.
The gravitational force $\mathbf{g}$ is directed in the negative
Z-direction:
\begin{equation}
\mathbf{g}= -g~\hat{z},
\end{equation}
where $g$ is the gravitational acceleration constant. In this notation,
the Rayleigh number may be defined as:
\begin{equation}
R= \frac{\alpha L^4 \mathbf{g}\cdot\boldsymbol{\nabla}T_0}{\nu D_T} =
- \frac{\alpha L^4 g \nabla T_0}{\nu D_T},
\end{equation}
where $\alpha$ is the thermal expansion coefficient, $\nu$ the
kinematic viscosity and $D_T$ the thermal diffusivity of the fluid.

We shall evaluate the structure factor of the fluid maintained in
a convection-free thermal nonequilibrium state, where the average
value $\langle \mathbf{v} \rangle$ of the fluid velocity
$\mathbf{v}$ will be zero. Such states correspond to both negative and
positive values of the Rayleigh number $R$ as long as $R$ is smaller
than the critical value $R_\mathrm{c}$. For this purpose we write the
linearized fluctuating Boussinesq equations in the form~\cite{SwiftHohenberg,CrossHohenberg}:
\begin{subequations}
\label{RB}
\begin{equation}
\frac{\partial}{\partial t}
\left(\nabla^2 w \right)=\nu~\nabla^2 \left(\nabla^2 w \right) +
\alpha g \left(\frac{\partial^2 \theta}{\partial x^2}+\frac{\partial^2 \theta}{\partial y^2}\right)
+ F_1,
\label{RB:1}
\end{equation}
\begin{equation}
\frac{\partial\theta}{\partial t}=
D_T~\nabla^2\theta - w~\nabla T_0 + F_2
\label{RB:2}
\end{equation}
\end{subequations}
where $\theta=T-T_0$ represents the local fluctuating temperature and
$w$ is the fluctuating Z-component of the fluid velocity $\mathbf v$.
To eliminate the stationary pressure gradient from the equations we
find it convenient to consider Eq.~(\ref{RB:1}) for $\nabla^2 w$,
rather than an equation for the fluctuating fluid velocity
$\mathbf{v}$ itself~\cite{Chandra}. Finally, $F_1$ and $F_2$ represent
the contributions from rapidly varying short-range fluctuations and
are related to Landau's random stress tensor $\delta\mathsf{T}$ and
random heat flow $\delta\mathbf{Q}$ in such a way that~\cite{HohenbergSwift}:
\begin{subequations}
\label{DefF}
\begin{equation}
F_1 = \frac{1}{\rho} \left\{\boldsymbol{\nabla}\times\left[\boldsymbol{\nabla}\times\left(\boldsymbol{\nabla}\cdot \delta{\mathsf T}\right)\right]\right\}_z
\label{DefF:a}
\end{equation}
\begin{equation}
F_2 = -\frac{D_T}{\lambda} \boldsymbol{\nabla}\left(\delta{\bf Q}\right),
\label{DefF:b}
\end{equation}
\end{subequations}
where $\rho$ and $\lambda$ are the density and the thermal conductivity
of the fluid, while the subscript $z$ in Eq.~(\ref{DefF:a}) indicates
that $F_1$ has to be identified with the Z-component of the vector between
the curly brackets. We note that we are here considering only the effects
of additive noise represented by $F_1$ and $F_2$ in Eqs.~(\ref{RB}).
Multiplicative noise has been also considered in the
literature~\cite{GarciaOjalvoEtAl}, but that subject lies outside the scope
of the present paper.
Since $\langle\theta\rangle=0$ and $\langle w\rangle =0$
everywhere in the fluid layer, we note that the solution of the fluctuating
Boussinesq equations will be independent of the sign adopted for
$\theta$ or $w$~\cite{SwiftHohenberg,CrossHohenberg,Chandra,HohenbergSwift}.

The conditions for the validity of the Boussinesq equations, also referred
to as Oberbeck-Boussinesq equations, have been discussed extensively
in the literature~\cite{GrayGiorgini,Tritton,NormandPomeauVelarde}.
Specifically, in deriving the Boussinesq equations from the more general
hydrodynamic equations, one neglects the temperature dependence of the
various thermophysical properties, except for the density which is taken to
vary linearly with temperature. Deviations of this so-called Boussinesq
approximation do affect the characteristic nature of pattern formation
upon the appearance of convection~\cite{CrossHohenberg,Busse,Ahlers,BodenschatzEtAl2}.
However, the intensity of the nonequilibrium fluctuations in 
convection-free states below the convective instability appears to be much
less sensitive to deviations from the Boussinesq approximation.
Segr\`{e} et al. have analyzed and measured the temperature fluctuations
in liquid toluene subjected to temperature gradients up to 220 K
cm$^{-1}$~\cite{SegreEtAl1}. At the higher value of the temperature
gradient some of the thermophysical properties varied significantly
over the height of the fluid layer. Nevertheless, the intensity of the
observed nonequilibrium temperature fluctuations was equal to the intensity
predicted when all thermophysical-property values are taken at their average value
in the fluid layer. Similar results were obtained by Ahlers, measuring
the critical Rayleigh number for the onset of Rayleigh-B\'{e}nard
convection and the heat transfer slightly above the
threshold~\cite{Ahlers}.

We note that in the derivation of the Boussinesq equations it is also
assumed that the adiabatic temperature gradient $(\alpha \bar{T}_0/c_P)g$,
where $c_P$ is the isobaric specific heat capacity and $\bar{T}_0$ the
average temperature in the fluid layer, is small and can be neglected
compared to the magnitude of the imposed temperature gradient.
In practice, this is a very good approximation~\cite{GrayGiorgini,Tritton}.
In addition, in the Boussinesq Eq.~(\ref{RB:2}) the coefficient multiplying
$\nabla^2 \theta$ is usually identified with the thermal diffusivity of
the fluid $D_T$. For consistency, we have also expressed 
the prefactor of the second random noise term in Eq.~(\ref{DefF:b}) in terms
of the same diffusivity $D_T$~\cite{Tritton}.

\section{Bulk structure factor of a fluid subjected to a stationary
temperature gradient}
\label{sec:1}
Starting from the fluctuating Boussinesq equations, we first derive an
expression for the structure factor of a fluid subjected to a stationary
temperature gradient in the presence of gravity but in the absence of
any boundary conditions. Althought the resulting expression for such
a bulk structure factor will only be meaningful for negative Rayleigh
numbers, it will enable us to develop a procedure most suited for
subsequent incorporation of boundary effects.

In the absence of any boundary conditions, a temporal and spatial
Fourier transformation can be applied to Eqs.~(\ref{RB}) to obtain
a set of equations for the fluctuations in the vertical component
of the velocity $w(\omega,\mathbf q)$ and for the fluctuations in
the temperature $\theta(\omega,\mathbf q)$ as a function of the
frequency $\omega$ and the wavevector $\mathbf q$:
\begin{equation}
\begin{pmatrix}
- q^2(\mathrm{i}~\omega+\nu~q^2)& \alpha~g~q_\parallel^2 \\
\nabla T_0 & \mathrm{i}~\omega + D_T~q^2
\end{pmatrix}
\begin{pmatrix}
w(\omega,{\mathbf q})  \\
\theta (\omega,{\mathbf q})
\end{pmatrix} =
\begin{pmatrix}
F_1 (\omega,{\mathbf q}) \\
F_2 (\omega,{\mathbf q})  
\end{pmatrix}, 
\label{Bulk1}
\end{equation}
where $q_\parallel = \sqrt{q_x^2+q_y^2}$ represents the
magnitude of the component of the wavevector $\mathbf{q}$ in
the XY-plane, i.e., the component of ${\mathbf q}$ perpendicular to the
temperature gradient. We note that in this paper the parallel and perpendicular
directions are defined with respect to the horizontal plane, as done by most
researchers in the field~\cite{SchmitzCohen1,KirkpatrickCohen,SchmitzCohen2,VanBeijerenCohen},
but unlike the notation
used by Segr\`{e} et al.~\cite{SegreSchmitzSengers,SegreSengers}.
The Fourier transforms $F_1(\omega,\mathbf q)$ and $F_2(\omega,\mathbf q)$
of the random noise terms can be expressed as a function of the Fourier
transforms of the random stress tensor and the random heat flow:
\begin{subequations}
\begin{equation}
F_1 (\omega,{\mathbf q})=
\frac{\mathrm i}{\rho} \left\{q^2\left[{\mathbf q}\cdot\delta{\mathsf T}(\omega,{\mathbf q})\right]_z -
q_z {\mathbf q}\cdot\left[{\mathbf q}\cdot\delta{\mathsf T}(\omega,{\mathbf q})\right]\right\},
\end{equation}
\begin{equation}
F_2 (\omega,{\mathbf q})= - {\mathrm i}~
\frac{D_T}{\lambda} {\mathbf q}\cdot\delta{\mathbf Q}(\omega,{\mathbf q}).
\end{equation}
\end{subequations} 
The solution for $w(\omega,\mathbf q)$
and $\theta(\omega,\mathbf q)$ can be readily obtained by inverting
the matrix in Eq.~(\ref{Bulk1}), so that:
\begin{equation}
\begin{pmatrix}
w(\omega,{\mathbf q})  \\
\theta (\omega,{\mathbf q})
\end{pmatrix} =
\frac{
\begin{pmatrix}
\mathrm{i}~\omega + D_T~q^2 & -\alpha g q_\parallel^2 \\
-\nabla T_0 &- q^2\left(\mathrm{i}~\omega+\nu~q^2\right)
\end{pmatrix}
}{q^2~\left[\omega - \mathrm{i} \Omega_+(\mathbf{q})\right]~\left[\omega - \mathrm{i} \Omega_-(\mathbf{q})\right]}
\begin{pmatrix}
F_1 (\omega,{\mathbf q}) \\
F_2 (\omega,{\mathbf q})  
\end{pmatrix},
\label{MatrizInversa}
\end{equation}
where we have introduced the quantities:
\begin{equation}
\Omega_\pm(\mathbf q) = \frac{1}{2}~q^2~
\left[ (\nu+D_T) \pm \sqrt{(\nu-D_T)^2- 4g\alpha\nabla T_0\frac{q_\parallel^2}{q^6}} \right].
\label{DefOmega}
\end{equation}
Stricly speaking, in the expression~(\ref{DefOmega}) for the decay rates
$\Omega_\pm(\mathbf q)$ the temperature gradient $\nabla T_0$ should
be identified with the effective temperature gradient
$\nabla T_0 +(\alpha \bar{T}_0/c_P)g$, as shown by Segr\`{e} et al.~\cite{SegreSchmitzSengers}.
However, as mentioned in Section~\ref{sec:0}, the contribution
$(\alpha \bar{T}_0/c_P)g$ from the adiabatic temperature gradient is neglected
in the Boussinesq approximation. In the limit $g\to 0$, we have
$\Omega_+ = \nu q^2$ and $\Omega_-= D_T q^2$, which are the decay rates associated
with the transverse-velocity fluctuations $w$ and the temperature
fluctuations $\theta$, respectively.

In the Boussinesq approximation, density fluctuations are caused only
by temperature fluctuations, while pressure fluctuations are neglected.
Consequently, the structure factor $S(\omega,{\textbf q})$ will be
related to the autocorrelation function of the temperature fluctuations:
\begin{equation}
\langle \theta^{*}(\omega,{\mathbf q})~
\theta(\omega^\prime,{\mathbf q}^\prime)\rangle=
\frac{1}{\alpha^2\rho^2}~S(\omega,{\mathbf q})~(2 \pi)^4~
\delta(\omega-\omega^\prime)~
\delta({{\mathbf q}-{\mathbf q}^\prime}).
\label{DefS}
\end{equation}
To deduce the autocorrelation function
$\langle \theta^{*}(\omega,{\mathbf q})~
\theta(\omega^\prime,{\mathbf q}^\prime)\rangle$
of the temperature fluctuations from Eqs.~(\ref{MatrizInversa}), we need
the correlation functions for the Langevin noise terms
$F_1(\omega,\mathbf{q})$ and $F_2(\omega,\mathbf{q})$,
defined by Eqs.~(\ref{DefF}).
In nonequilibrium fluctuating hydrodynamics it is
assumed that the random stress tensor and the random heat flow
retain their local-equilibrium values~\cite{SchmitzCohen1,LandauLifshitz,RonisEtAl}.
Thus, as in equilibrium, we shall consider both $\delta{\mathsf T}$ as
$\delta{\mathbf Q}$ as ``white" noise. Their autocorrelation functions are short ranged
in time and in space, being proportional to delta functions.
Fourier transforming the explicit expressions for the correlation functions
of $\delta{\mathsf T}(\mathbf{r},t)$ and
$\delta{\mathbf Q}(\mathbf{r},t)$ as, for instance,
given by Eqs. (3.12) in Ref.~\cite{SchmitzCohen1}, and using the
definition~(\ref{DefF}) of $F_1(\omega,{\mathbf q})$ and
$F_2(\omega,{\mathbf q})$, we obtain:
\begin{eqnarray}
\langle F_1^{*}(\omega,{\mathbf q})~
F_1(\omega^\prime,{\mathbf q}^\prime)\rangle&=&
2 k_\mathrm{B} T \frac{\nu}{\rho}~q_\parallel^2~q^4~
(2 \pi)^4~
\delta(\omega-\omega^\prime)~
\delta({{\mathbf q}-{\mathbf q}^\prime}),
\nonumber \\
\langle F_2^{*}(\omega,{\mathbf q})~
F_2(\omega^\prime,{\mathbf q}^\prime)\rangle&=&
\frac{2 k_\mathrm{B} T^2\lambda}{\rho^2 c_P^2}~q^2~
(2 \pi)^4~
\delta(\omega-\omega^\prime)~
\delta({{\mathbf q}-{\mathbf q}^\prime}),
\label{CorrF} \\
\langle F_1^{*}(\omega,{\mathbf q})~
F_2(\omega^\prime,{\mathbf q}^\prime)\rangle&=&
\langle F_2^{*}(\omega,{\mathbf q})~
F_1(\omega^\prime,{\mathbf q}^\prime)\rangle = 0, \nonumber
\end{eqnarray}
where $k_\mathrm{B}$ represents Boltzmann's constant.
Combining Eqs.~(\ref{MatrizInversa}), (\ref{DefS}) and
(\ref{CorrF}) we obtain for the dynamic structure factor
$S(\omega,{\mathbf q})$ of a nonequilibrium fluid in the Boussinesq
approximation the following expression:
\begin{equation}
S(\omega,{\mathbf q})=
S_\mathrm{E}
\left\{\frac{2 D_T q^2 \left(\omega^2+\nu^2 q^4\right)}
{\left[\omega^2+\Omega_+^2(\mathbf{q})\right]~\left[\omega^2+\Omega_-^2(\mathbf{q})\right]}
+ \frac{2 \nu (c_P/T)~\nabla T_0^2 q_\parallel^2}
{\left[\omega^2+\Omega_+^2(\mathbf{q})\right]~\left[\omega^2+\Omega_-^2(\mathbf{q})\right]}
\right\},
\label{Bulk2}
\end{equation}
where $S_\mathrm{E}$ represents the intensity of the fluctuations
in thermodynamic equilibrium:
\begin{equation}
S_\mathrm{E} =  \rho^2 \kappa_T k_\mathrm{B} T \frac{\gamma -1}{\gamma}.
\label{SE}
\end{equation}
Here $\gamma$ denotes the heat-capacity ratio $c_P/c_V$ and
$\kappa_T$ is the isothermal compressibility. In deriving
Eq.~(\ref{Bulk2}) we have employed the thermodynamic
relation $\alpha^2 D_T = [(\gamma -1)/\gamma] \lambda \kappa_T / T$.

The equation for the Rayleigh spectrum of a fluid subjected to a stationary
temperature gradient including the contribution from gravity was
first derived by Segr\`{e} et al.~\cite{SegreSchmitzSengers}. We recover
the same result, but without the (negligible) contribution from the
adiabatic temperature gradient. In the limit $g\to 0$, Eq.~(\ref{Bulk2})
reduces to the expression of the Rayleigh spectrum of a fluid
subjected to a stationary temperature gradient, first obtained by
Kirkpatrick et al.~\cite{KirkpatrickEtAl}, and subsequently reproduced
by other investigators~\cite{RonisProcaccia,SchmitzCohen1,LawSengers}.
The nonequilibrium structure factor, as given by Eq.~(\ref{Bulk2}),
is anisotropic, and it depends on the magnitude of the horizontal
component $q_\parallel$ of the scattering vector directly and through
the decay rates $\Omega_\pm(\mathbf{q})$.

In this paper we shall focus our attention on the static structure
factor:
\begin{equation}
S({\mathbf q}) = (2\pi)^{-1} \int d\omega~S(\omega,{\mathbf q}),
\label{DefStatic}
\end{equation}
which determines the total intensity of Rayleigh scattering~\cite{BernePecora}.
For subsequent use, we find it convenient to introduce dimensionless
wave numbers $\tilde{q}=qL$ and $\tilde{q}_\parallel=q_\parallel L$,
where $L$ is the finite height of the fluid layer.
Upon substituting Eq.~(\ref{DefOmega}) for $\Omega_\pm(\mathbf{q})$ into
Eq.~(\ref{Bulk2}) and integrating the resulting expression for
$S(\omega,\mathbf{q})$, we obtain:
\begin{equation}
S({\mathbf q})= S_\mathrm{E}~
\left\{1 + \tilde{S}^0_\mathrm{NE}~
\cfrac{\tilde{q}_\parallel^2}{(\tilde{q}^6- R ~\tilde{q}_\parallel^2)}
\right\}.
\label{Bulk3}
\end{equation}
In Eq.~(\ref{Bulk3}) $\tilde{S}^0_\mathrm{NE}$
represents the strength of the nonequilibrium
enhancement of the structure factor, which is given by:
\begin{equation}
\tilde{S}^0_\mathrm{NE}= \frac{\sigma R}{\sigma+1}+
\frac{(\sigma-1) (c_P/T) L^4}{\nu^2-D_T^2}~(\nabla T_0)^2,
\label{strength}
\end{equation}
where $\sigma=\nu/D_T$ is the Prandtl number. Since we have
not yet considered any finite-size effects, Eqs.~(\ref{Bulk3})
and~(\ref{strength}) do not depend explicitly on the
height $L$. Equations~(\ref{Bulk3}) and~(\ref{strength}) are
identical to Eq.~(2.35) in Ref.~\cite{SegreSchmitzSengers}.
The term in Eq.~(\ref{strength}) proportional to $R$ is related to the
adiabatic
temperature gradient and is in practice negligibly small. Hence,
the intensity of nonequilibrium fluctuations continues to be
proportional to $(\nabla T_0)^2$ in the presence of gravity.

Note that~(\ref{Bulk3}) can only be valid for negative Rayleigh
number $R$, i.e., when the fluid layer is heated from above so that
$\nabla T_0>0$. For any $R>0$, the nonequilibrium contribution to the
structure factor will always diverge at some finite value of the
horizontal component $q_\parallel$ of the scattering vector.
The appearance of this divergence is a direct consequence of us
having performed the calculation without taking into account boundary
conditions. As will be demonstrated in the next section, when boundary
conditions are taken into account, a different expression for
$S({\mathbf q})$ is obtained. This new expression will be valid not only
for negative Rayleigh numbers, but also for a finite interval of
positive Rayleigh numbers, up to some critical Rayleigh number
$R_\mathrm{c}$.

\section{Modification of the nonequilibrium structure factor due
to finite-size effects}
\label{sec:2}
Since in practice the fluid layer is confined between two horizontal
plates separated by a (small) distance $L$, the nonequilibrium structure
factor will be affected by the presence of boundary conditions in the
Z-direction. In some previous publications, we have evaluated finite-size
effects on the nonequilibrium structure factor neglecting
gravity~\cite{Physica,EPJ,Mexico}. A major conclusion of this previous
work is that finite-size effects appear at wave numbers where in practice
the nonequilibrium structure factor will also be affected by gravity.
Hence, for the interpretation of experimental measurements it becomes
imperative to incorporate both finite-size effects and gravity effects
simultaneously. This task is implemented in the present section.

As in our previous publication~\cite{Physica}, we again apply a Fourier
transformation
of the fluctuating Boussinesq equations~(\ref{RB}) in space and in
time, but restricting the spatial Fourier transformation to the XY-plane.
We thus obtain the following set of linear stochastic differential
equations:
\begin{equation}
\begin{pmatrix}
\mathrm{i}~\omega~\left[\cfrac{d^2}{dz^2}-q_\parallel^2\right] -
\nu~\left[\cfrac{d^2}{dz^2}-q_\parallel^2\right]^2& \alpha~g~q_\parallel^2 \\
\nabla T_0 & \mathrm{i}~\omega - D_T~\left[\cfrac{d^2}{dz^2}-q_\parallel^2\right]
\end{pmatrix}
\begin{pmatrix}
w(\omega,{\mathbf q_\parallel},z)  \\
\theta (\omega,{\mathbf q_\parallel},z)
\end{pmatrix} =
\begin{pmatrix}
F_1 (\omega,{\mathbf q_\parallel},z) \\
F_2 (\omega,{\mathbf q_\parallel},z)  
\end{pmatrix},
\label{Finite1}
\end{equation}
where $\mathbf{q}_\parallel$ is the wavevector in the XY-plane.
The random noise terms $F_1(\omega,{\mathbf q}_\parallel,z)$ and
$F_2(\omega,{\mathbf q}_\parallel,z)$ in Eqs.~(\ref{Finite1}) are
related to the partial Fourier transforms
$\delta{\mathsf T}(\omega,{\mathbf q}_\parallel,z)$ of the random
stress tensor and $\delta{\mathbf Q}(\omega,{\mathbf q}_\parallel,z)$
of the random heat flux. The actual expressions are a bit complicated and
can be found elsewhere~\cite{EPJ}.

As is often done in the literature~\cite{KirkpatrickCohen,Chandra}, for the sake
of simplicity, we adopt stress-free boundary conditions for the
vertical velocity and perfectly conducting walls for the
temperature, so that:
\begin{equation}
\begin{array}{rcccc}
\theta(\omega,\mathbf{q}_\parallel,z)&=&0&\hspace{1 cm}{\rm at}\hspace{1 cm}&z=0,L~, \\
w(\omega,\mathbf{q}_\parallel,z)&=&0&\hspace{1 cm}{\rm at}\hspace{1 cm}&z=0,L~, \\
\dfrac{d^2}{dz^2}~w(\omega,\mathbf{q}_\parallel,z)&=&0&\hspace{1 cm}{\rm at}\hspace{1 cm}&z=0,L~. 
\end{array}
\label{SF}
\end{equation}
Note that these boundary conditions imply the absence of any
possible fluctuations in the temperature and velocity of the
fluid adjacent to the walls. As is well known, stress-free
boundary conditions represent a fluid bounded by two free
surfaces, which is a rather unrealistic case~\cite{Chandra,Manneville}.
For the realistic case of a fluid confined between two rigid
surfaces, no-slip boundary conditions in the velocity are 
more appropriate. For the particular case of $g=0$,
we have evaluated the nonequilibrium structure factor using both
stress-free~\cite{Physica} and no-slip~\cite{EPJ} boundary
conditions. While there are numerical differences between the two
solutions, the dependence of the nonequilibrium structure factor
on the wave number $q$ appears to be qualitatively the same for
the two sets of boundary conditions. For mathematical simplicity,
we evaluate here the nonequilibrium structure factor using
stress-free boundary conditions in the expectation that also in the
presence of gravity the more realistic no-slip boundary conditions
will not modify the qualitative nature of the wave-number dependence
of the structure factor.

To search for a solution of Eq.~(\ref{Finite1}), we represent
$w(\omega,{\mathbf q}_\parallel,z)$ and $\theta(\omega,{\mathbf q}_\parallel,z)$
as a series expansion in a complete set of eigenfunctions of the
differential operator in Eq.~(\ref{Finite1}), satisfying the
boundary conditions~(\ref{SF})~\cite{Physica}. Because of the simplicity of the
boundary conditions considered here, an appropriate set of eigenfunctions is
the Fourier sine basis in the $[0,L]$ interval~\cite{Chandra}.
We thus assume:
\begin{equation}
\begin{pmatrix}
w(\omega,\mathbf{q}_{\parallel},z)\\
\theta(\omega,\mathbf{q}_{\parallel},z)
\end{pmatrix}=
\sum_{N=1}^{\infty}
\begin{pmatrix}
A_N(\omega,{\mathbf q}_{\parallel})\\
B_N(\omega,{\mathbf q}_{\parallel}) \end{pmatrix}
\sin\genfrac{(}{)}{}{}{N\pi z}{L}.
\label{P10}
\end{equation}

To deduce the coefficients
$A_N(\omega,\mathbf{q}_\parallel)$ and $B_N(\omega,\mathbf{q}_\parallel)$
from Eq.~(\ref{Finite1}),
we also have to represent the random noise terms
$F_1(\omega,{\mathbf q}_{\parallel},z)$ and
$F_2(\omega,{\mathbf q}_{\parallel},z)$
as a Fourier sine series:
\begin{equation}
\begin{pmatrix}
F_1(\omega,{\mathbf q}_{\parallel},z)\\
F_2(\omega,{\mathbf q}_{\parallel},z)
\end{pmatrix}=
\sum_{N=1}^{\infty}
\begin{pmatrix}
F_{1,N}(\omega,\mathbf{q}_{\parallel})\\
F_{2,N}(\omega,\mathbf{q}_{\parallel})
\end{pmatrix}
\sin\genfrac{(}{)}{}{}{N\pi z}{L},
\label{P11}
\end{equation}
where we have introduced the set of random functions
$F_{1,N}(\omega,{\mathbf q}_\parallel)$ and
$F_{2,N}(\omega,{\mathbf q}_\parallel)$, which are the projections
of the random noise terms over the eigenfunction basis.
They are given by:
\begin{equation}
\begin{pmatrix}
F_{1,N}(\omega,\mathbf{q}_{\parallel})\\
F_{2,N}(\omega,\mathbf{q}_{\parallel}) \end{pmatrix}=
{2\over L} \int_0^L 
\begin{pmatrix}
F_1(\omega,{\mathbf q}_{\parallel},z)\\
F_2(\omega,{\mathbf q}_{\parallel},z) \end{pmatrix}
\sin\genfrac{(}{)}{}{}{N\pi z}{L} dz.
\label{P12}
\end{equation}
Representing the random noise terms by Eq.~(\ref{P11}), one readily
deduces from Eq.~(\ref{Finite1}) expressions for the coefficients of the
Fourier series $A_N(\omega,\mathbf{q}_\parallel)$ and
$B_N(\omega,\mathbf{q}_\parallel)$. To calculate the structure factor in the
Boussinesq approximation, we only need the
coefficients $B_N(\omega,\mathbf{q}_\parallel)$ for the temperature fluctuations.
Substituting Eqs.~(\ref{P10}) and (\ref{P11}) into Eq.~(\ref{Finite1}) we obtain:
\begin{equation}
B_N(\omega,\mathbf{q}_\parallel)=\frac{-\nabla T_0 F_{1,N}(\omega,\mathbf{q}_\parallel)-
\left(\frac{N^2\pi^2}{L^2}+q_\parallel^2\right)
\left[\mathrm{i}~\omega+\nu\left(\frac{N^2\pi^2}{L^2}+q_\parallel^2\right)\right]
F_{2,N}(\omega,\mathbf{q}_\parallel)}
{\left(\frac{N^2\pi^2}{L^2}+q_\parallel^2\right)
\left(\omega - \mathrm{i}~\Omega_+(N,q_\parallel)\right)
\left(\omega - \mathrm{i}~\Omega_-(N,q_\parallel)\right)},
\end{equation}
where, similarly to Eq.~(\ref{DefOmega}), we have introduced
decay rates $\Omega_\pm(N,q_\parallel)$, which are now given by:
\begin{equation}
\Omega_\pm(N,q_\parallel)=
\frac{1}{2} \left(\tfrac{N^2\pi^2}{L^2}+q_\parallel^2\right)
\left[ (\nu+D_T) \pm \sqrt{(\nu-D_T)^2 - \frac{4g\alpha\nabla T_0~q_\parallel^2}{\left(\frac{N^2\pi^2}{L^2}+q_\parallel^2\right)^3}} \right].
\end{equation}

Since the differential operator in Eq.~(\ref{Finite1}) depends only
on the modulus of $\mathbf{q}_\parallel$, the problem has cylindrical
symmetry and the solution will only depend on the magnitude
$q_\parallel$ of the two-dimensional wavevector $\mathbf{q}_\parallel$.

In analogy to Eq.~(\ref{DefS}), the relationship between the dynamic
structure factor $S(\omega,q_\parallel,z,z^\prime)$ and the autocorrelation
function of the temperature fluctuations is now given by:
\begin{equation}
\langle \theta^{*}(\omega,{\mathbf q}_\parallel,z)~
\theta(\omega^\prime,{\mathbf q}_\parallel^\prime,z^\prime)\rangle =
\frac{(2 \pi)^3}{\alpha^2\rho^2}~S(\omega,q_\parallel,z,z^\prime) 
~\delta(\omega-\omega^\prime)~
\delta({{\mathbf q}_\parallel-{\mathbf q}_\parallel^\prime})
\label{DefS2}
\end{equation}

To obtain the structure factor $S(\omega,q_\parallel,z,z^\prime)$
we need the correlation functions between the different projections
of the random noise terms. These have been calculated
in a previous publication with the result~\cite{Physica}:
\begin{eqnarray}
\langle F^*_{1,N}(\omega,{\mathbf q}_{\parallel}) \cdot F_{1,M}(\omega^\prime,{\mathbf q}_{\parallel}^\prime)  \rangle &=&
2 k_\mathrm{B} T {\nu\over\rho}{2\over L}~q_{\parallel}^2 \left(q_\parallel^2+{N^2\pi^2\over L^2}\right)^2~\delta_{NM} \nonumber \\
&\times&(2\pi)^3~\delta(\omega-\omega^\prime)~\delta({\mathbf q}_{\parallel}-{\mathbf q}_{\parallel}^\prime), \nonumber\\
\langle F^*_{1,N}(\omega,{\mathbf q}_{\parallel}) \cdot F_{2,M}(\omega^\prime,{\mathbf q}_{\parallel}^\prime)  \rangle &=&
\langle F^*_{2,N}(\omega,{\mathbf q}_{\parallel}) \cdot F_{1,M}(\omega^\prime,{\mathbf q}_{\parallel}^\prime)  \rangle = 0, \label{corr2}\\
\langle F^*_{2,N}(\omega,{\mathbf q}_{\parallel}) \cdot F_{2,M}(\omega^\prime,{\mathbf q}_{\parallel}^\prime)  \rangle &=&
{2 k_\mathrm{B} T^2 \lambda\over\rho^2 c_P^2} {2\over L}~\left(q_\parallel^2+{N^2\pi^2\over L^2}\right)~\delta_{NM} \nonumber \\
&\times&(2\pi)^3~\delta(\omega-\omega^\prime)~\delta({\mathbf q}_{\parallel}-{\mathbf q}_{\parallel}^\prime). \nonumber
\end{eqnarray}
As in the case of the absence of any boundary conditions, we continue
to assume that the correlation functions between the different components
of the random current tensor and the random heat flux retain their
equilibrium values. This assumption remains valid as long as
$L$ is a macroscopic distance, much larger than the molecular
distances in the fluid.

In this paper we are interested in the static structure factor
$S(q_\parallel,z,z^\prime)=(2\pi)^{-1} \int d\omega S(\omega,q_\parallel,z,z^\prime)$.
With the information presented above, this quantity can be readily
calculated. In analogy to Eq.~(\ref{Bulk3}) for $S(\mathbf{q})$ in the
absence of any boundary conditions, the modified nonequilibrium
structure factor can be written in the form:
\begin{equation}
S(q_\parallel,z,z^\prime)=S_\mathrm{E}~
\left[\delta(z-z^\prime)+\tilde{S}^0_\mathrm{NE}~
\tilde{S}_\mathrm{NE}(q_\parallel,z,z^\prime)\right],
\label{G23}
\end{equation}
where $S_\mathrm{E}$ and $\tilde{S}^0_\mathrm{NE}$
are again given by Eqs.~(\ref{SE}) and~(\ref{strength}), respectively,
while $\tilde{S}_\mathrm{NE}(q_\parallel,z,z^\prime)$ incorporates
the finite-size effects and is given by:
\begin{equation}
\tilde{S}_\mathrm{NE}(q_\parallel,z,z^\prime)=
\frac{2 \tilde{q}_\parallel^2}{L} \sum_{N=1}^{\infty}
\frac{\sin\left(\dfrac{N \pi z}{L}\right)
\sin\left(\dfrac{N \pi z^\prime}{L}\right)}
{(\tilde{q}_\parallel^2 + N^2 \pi^2)^3-R~\tilde{q}_\parallel^2}.
\label{G24}
\end{equation}

The first term in Eq.~(\ref{G23}) is again the static structure factor
of a fluid in thermodynamic equilibrium, which is not affected by
any finite-size effects~\cite{BernePecora}. The second term in Eq.~(\ref{G23})
represents the nonequilibrium enhancement of the structure factor.
This nonequilibrium enhancement is proportional to $(\nabla T_0)^2$
through the expression~(\ref{strength}) for $\tilde{S}_\mathrm{NE}^0$,
it depends on the gravitational acceleration constant $g$ through
the appearance of the Rayleigh number in Eqs.~(\ref{strength})
and~(\ref{G23}), and it depends on the finite height $L$ of the fluid
layer explicitly in Eq.~(\ref{G23}) and also through $\tilde{q}_\parallel=q_\parallel L$.
In the limit $R\to 0$ we recover the expression
for $S(q_\parallel,z,z^\prime)$ presented in a previous
publication~\cite{Physica}.

It is important to note that Eq.~(\ref{G24}) is valid for both
negative and positive Rayleigh numbers, provided that
\begin{equation}
R<R_\mathrm{c} = \frac{27 \pi^4}{4}.
\label{Rcrit}
\end{equation}
For $R>R_\mathrm{c}$ there always exist values of $q_\parallel$
for which right-hand side of Eq.~(\ref{G24}) diverges.
Of course, the value $R_\mathrm{c} = 27 \pi^4/4$ equals the well-known
value obtained form a linear stability analysis of the Boussinesq
equations with no-slip boundary conditions~\cite{Chandra}.
As expected, for $R\geq R_\mathrm{c}$ the present calculation breaks
down because the fluctuations around the conductive state are no
longer small.

Equation~(\ref{G24}) for the normalized nonequilibrium enhancement
$\tilde{S}_\mathrm{NE}(\tilde{q}_\parallel,z,z^\prime)$ represents
our principal result for the combined effects of gravity and finite-size
on the nonequilibrium structure factor of the fluid. The remaining
part of this paper is concerned with an analysis of some of the
physical consequences that follow from Eq.~(\ref{G24}).

The double Fourier series contained in Eq.~(\ref{G24}), when it
is convergent, can be easily summed. To do so, we first find the
roots of the denominator for $N^2$ . We then split the inverse of the
product of three binomials as a sum of three terms, and we finally obtain
for $\tilde{S}_\mathrm{NE}(\tilde{q}_\parallel,z,z^\prime)$:
\begin{equation}
\tilde{S}_\mathrm{NE}(\tilde{q}_\parallel,z,z^\prime)=\frac{2}{3 L} \tilde{q}_\parallel^{2/3} 
~\sum_{N=1}^{\infty}\sum_{i=0}^{2} \dfrac{\sin{\left(\dfrac{N \pi z}{L}\right)}\sin{\left(\dfrac{N \pi z^\prime}{L}\right)}}
{\lambda_i^2 \left[N^2 \pi^2 - \left(\dfrac{\lambda_i}{\tilde{q}_\parallel^{4/3}}-1\right)\tilde{q}_\parallel^2\right]},
\label{G25}
\end{equation}
where $\lambda_i$ for $i=0,1,2$ are the three complex cubic
roots of $R$:
\begin{equation}
\lambda =
|R|^{1/3}
\begin{pmatrix}
\exp{\mathrm{i}~\frac{\psi}{3}} \\
\exp{\mathrm{i}~\frac{\psi +2\pi}{3}} \\
\exp{\mathrm{i}~\frac{\psi +4\pi}{3}} 
\end{pmatrix}.
\label{DefLa}
\end{equation}
In Eq.~(\ref{DefLa}) $\psi$ represents the phase of $R$ when considered as
a complex number: $\psi=0$ for $R>0$ and $\psi=\pi$ for $R<0$. This procedure
is only valid for $R\neq 0$. However, the finite-size effects on the
structure factor for the particular case $R=0$ have already been considered
in a previous publication~\cite{Physica}, where the sum of the series
in Eq.~(\ref{G24}) for this particular case has been performed by a different
method. Hence, we do not need to re-evaluate $\tilde{S}_\mathrm{NE}$
for the case $R=0$. To implement the summation in Eq.~(\ref{G25}) we
consider the relation:
\begin{align}
&\sum\limits_{N=1}^{\infty} \frac{1}{N^2-\mu}
\sin\left(\dfrac{N \pi z}{L}\right)\sin\left(\dfrac{N \pi z^\prime}{L}\right)
= \label{G26} \\
&\frac{-\pi}{4\sqrt{\mu}}~\dfrac{\cos{\left(\dfrac{\pi\sqrt{\mu}}{L} (L-|z-z^\prime|) \right)}-\cos{\left(\dfrac{\pi\sqrt{\mu}}{L} (L-|z+z^\prime|) \right)}}
{\sin{(\pi\sqrt{\mu})}}. \nonumber
\end{align}
This equation can be obtained from formula 1.445 in Ref.~\cite{Gradstein} and
it is valid for $z, z^\prime \in [0,L]$ and for any complex number $\mu$,
provided that in the right-hand side of~(\ref{G26}) the square root of
$\mu$ with positive real part is chosen. Now, the sum of the double Fourier
series contained in Eq.~(\ref{G24}) can be performed, but the result is a
long expression, not particularly informative, which can be
easily obtained by the reader.
Explicit expressions to be obtained upon further integration
of $\tilde{S}_\mathrm{NE}(\tilde{q},z,z^\prime)$ will
be presented in the following section.

\section{Consequences for light scattering and shadowgraph experiments}
\label{sec:LS}
\subsection{Nonequilibrium structure factor probed in experiments}
\begin{figure}
\begin{center}
\resizebox{0.60\textwidth}{!}{\includegraphics{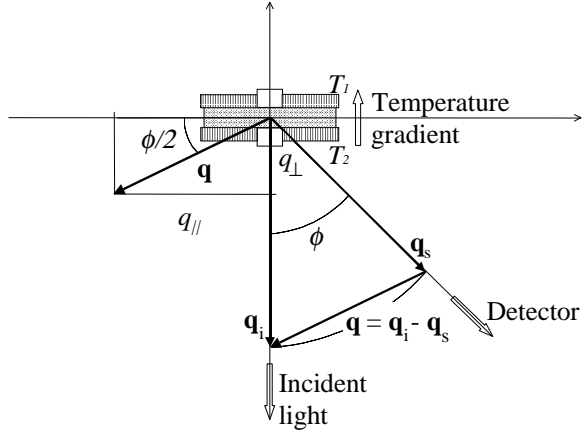}}
\end{center}
\caption{Schematic representation of a nonequilibrim light-scattering
experiment. $\mathbf{q}_\mathrm{i}$ is the wave vector of the incident
light and $\mathbf{q}_\mathrm{s}$ is the wave vector of the scattered light.
The magnitude $q=|\mathbf{q}_\mathrm{i}-\mathbf{q}_\mathrm{s}|$ of the
scattering wave vector is related to the scattering angle $\phi$ by
$q=2 q_0 \sin(\phi/2)$, where $q_0$ is the magnitude of the wave vector
$\mathbf{q}_\mathrm{i}$ of the incident light inside the fluid.}
\label{segunda}
\end{figure}

The nonequilibrium fluctuations can be obserbed in small-angle
light-scattering experiments. For a discussion of such light-scattering
experiments
we consider an experimental arrangement like the ones employed
at the University of Maryland~\cite{LawSengers,LawGammonSengers,LawEtAl,SegreEtAl1}
and by Vailati and Giglio at the University of
Milan~\cite{VailatiGiglio1,VailatiGiglio2}.
A schematic representation of such a light-scattering experiment
is shown in Fig.~\ref{segunda}. The scattering medium is a thin
horizontal fluid layer bounded by two parallel plates whose
temperatures can be controlled independently so as to establish
a temperature gradient across the fluid layer. The temperature
gradient can be parallel or antiparallel to the direction of gravity.
The horizontal plates
are furnished with windows allowing laser light to propagate through
the fluid in the direction parallel to the gravity and to the
temperature gradient. Light scattered over an angle $\phi$ arises
from fluctuations with a wave number such that~\cite{BernePecora}:
\begin{equation}
q = 2 q_0 \sin{(\phi /2)},
\label{P30}
\end{equation}
where $q_0$ is the wave number of the incident light inside the
scattering medium. To observe any nonequilibrium fluctuations one
needs to observe the scattered light at small wave numbers
and, hence, at small scattering angles.

From electromagnetic theory~\cite{BernePecora} it follows that
the scattering intensity  $S(\mathbf{q})$ is obtained from an
integration of the structure factor over the scattering volume,
so that~\cite{Physica,SchmitzCohen2}:
\begin{equation}
S(q_\parallel,q_\perp)=\frac{1}{L}
\int\limits_0^L \int\limits_0^L
{\text e}^{-i q_\perp (z-z^\prime)}~
S(q_\parallel,z,z^\prime)~dz~dz^\prime.
\label{P31}
\end{equation}
In Eq.~(\ref{P31}) we have assumed that the scattering volume extends
over the full height of the fluid layer as is the case in small-angle
light scattering from a thin fluid layer~\cite{SegreEtAl1}. In this
situation scattered
light received in the collecting pinhole of the detector indeed arises
from all points illuminated by the laser beam inside the fluid
layer. From Eq.~(\ref{P30}) and the geometrical arrangement shown in
Fig.~\ref{segunda}, we note that $q_\parallel$ and $q_\perp$ in an
actual light-scattering experiment are not independent variables,
because they are related to the scattering angle, $\phi$, by:
\begin{equation}
\begin{array}{lclcl}
q_\parallel &=&q \cos{(\phi/2)}&=& 2 q_0 \sin{(\phi/2)}\cos{(\phi/2)},\\
q_\perp &=&q \sin{(\phi/2)}&=& 2 q_0 \sin^2{(\phi/2)}.
\end{array}
\label{P32}
\end{equation}
Equation~(\ref{P32}) shows that for actual small-angle light-scattering
experiments one may use the approximation $q_\parallel \simeq q$,
$q_\perp\simeq 0$.

An alternative promising experimental technique for measuring
the intensity, though not the temporal spectrum, of nonequilibrium
fluctuations is quantitative shadowgraph
analysis~\cite{WuEtAl,DeBruynEtAl,BodenschatzEtAl,GiglioNature,BrogioliEtAl}.
The experimental arrangement is similar to the one depicted
in Fig.~\ref{segunda} but, instead of
a laser beam, an extended uniform monochromatic light source
is employed to illuminate the sample. Then many shadowgraph images
of a plane perpendicular to the temperature gradient are obtained
with a CCD detector, which measures the spatial distribution of intensity
$I(\mathbf{x})$, where $\mathbf{x}$ is a two-dimensional
position vector in the imaging plane. For each image
a shadowgraph signal is defined by:
\begin{equation}
i(\mathbf{x})=\frac{I(\mathbf{x})-I_0(\mathbf{x})}{I_0(\mathbf{x})},
\end{equation}
where $I_0(\mathbf{x})$ is the blank intensity distribution, when there
are no fluctuations in the index of refraction of the sample. In practice
$I_0(\mathbf{x})$ is calculated by averaging over many shadowgraph images,
so that fluctuations cancel out and the resulting $I_0(\mathbf{x})$
contains only contributions from nonuniform illumination
of the sample. From physical and geometrical optics one can 
demonstrate that the modulus squared of the two-dimensional Fourier
transform of the shadowgraph signal, $|i(\mathbf{q})|^2$, after taking an
azimuthal average, can be expressed as~\cite{DeBruynEtAl,BodenschatzEtAl,BrogioliEtAl}:
\begin{equation}
\overline{|i(\mathbf{q})|^2}(q) = \hat{T}(q)~S(q_\parallel=q,q_\perp =0),
\end{equation}
where the overline indicates the azimuthal average, in which case
the result depends only on the modulus $q_\parallel$ of the two-dimensional
Fourier vector $\mathbf{q}$. The factor $\hat{T}(q)$ represents an
optical transfer function, which can be derived from the optical arrangement
used to produce the shadowgraph pictures; it includes contributions
from the response of the CCD detector and the density dependence
of the refractive index~\cite{DeBruynEtAl,BodenschatzEtAl,BrogioliEtAl}.
Therefore, applying a two-dimensional Fourier transform to the shadowgraph
images, one can deduce the structure factor of the fluid
as a function of the wave number $q$ at $q_\perp =0$.

There exist an equivalence between small-angle
light scattering and shadow\-graphy, in the sense that both
methods give us $S(q_\parallel=q,q_\perp =0)$. For light
scattering, $q$ is the scattering wavevector as given by~(\ref{P30}),
whereas for shadowgraph techniques $q$ is the modulus of the
two-dimensional Fourier vector in the imaging plane. As mentioned
by Bodenschatz et al., what it is measured in the experiments
is a kind of vertical average of the fluctuations~\cite{BodenschatzEtAl}.

To obtain an explicit expression for the structure factor
$S(q_\parallel,q_\perp)$, we substitute Eq.~(\ref{G23}) with
$\tilde{S}_\mathrm{NE}(q_\parallel,z,z^\prime)$ given by Eq.~(\ref{G24})
into Eq.~(\ref{P31}), implement the summation by using the
relation~(\ref{G26}) and perform the double integration in
Eq.~(\ref{P31}). Introducing again dimensionless wave numbers
$\tilde{q}_\parallel=q_\parallel L$ and $\tilde{q}_\perp=q_\perp L$,
we thus obtain:
\begin{equation}
S(\tilde{q}_\parallel,\tilde{q}_\perp) =
S_\mathrm{E}
\left\{1 + \tilde{S}^0_\mathrm{NE}~
\tilde{S}_\mathrm{NE}(\tilde{q}_\parallel,\tilde{q}_\perp)\right\}
\label{G29}
\end{equation}
with
\begin{eqnarray}
\label{G30}
\tilde{S}_\mathrm{NE}(\tilde{q}_\parallel,\tilde{q}_\perp)&=&
\frac{\tilde{q}_\parallel^2}{\tilde{q}^6-R\tilde{q}_\parallel^2}
\\&+&
\frac{2\tilde{q}_\parallel}{3 R}
\sum\limits_{j=0}^{2} \frac{\lambda_j \sqrt{\lambda_j -\tilde{q}_\parallel^{4/3}}}
{\left[\tilde{q}^2-\lambda_j\tilde{q}_\parallel^{2/3}\right]^2}
~\frac{\cos{(\tilde{q}_\perp)}-\cos{\left(\tilde{q}_\parallel^{1/3} \sqrt{\lambda_j -\tilde{q}_\parallel^{4/3}}\right)}}
{\sin{\left(\tilde{q}_\parallel^{1/3} \sqrt{\lambda_j -\tilde{q}_\parallel^{4/3}}\right)}}.
\nonumber
\end{eqnarray}
In Eq.~(\ref{G29}), $S_\mathrm{E}$ and $S^0_\mathrm{NE}$
are again given by Eqs.~(\ref{SE}) and~(\ref{strength}), respectively,
while $\lambda_i$ in Eq.~(\ref{G30}) again
represent the three complex cubic roots of $R$ defined by Eq.~(\ref{DefLa}).
In thermodynamic equilibrium $\nabla T_0 = 0$ and $R=0$, so that
$S(\tilde{q}_\parallel,\tilde{q}_\perp)=S_\mathrm{E}$ and we recover from
Eq.~(\ref{G29}) the well-known isotropic equilibrium structure factor
given by Eq.~(\ref{SE}).

If we retain only the first term on the right-hand side of Eq.~(\ref{G30})
and substitute it in Eq.~(\ref{G29}) we recover the
expression~(\ref{Bulk3}) for the bulk structure factor of the fluid.
As noted in Section~\ref{sec:1}, this expression for the bulk structure
factor breaks down for any positive value of the Rayleigh number
because of a divergence at $q^6=q_\parallel^2 R L^{-4}$.
The second term in Eq.~(\ref{G30}) for
$\tilde{S}_\mathrm{NE}(\tilde{q}_\parallel,\tilde{q}_\perp)$,
containing a sum over the three complex cubic roots of $R$,
represents the finite-size effects on the nonequilibrium structure
factor and depends on the height $L$ of the fluid layer.
Although perhaps not immediately transparent, it can be verified that
this second term when multiplied by $\tilde{S}^0_\mathrm{NE}$ vanishes in the
limit $L \to \infty$. Moreover, for any finite value of $L$, the
divergence of the first term in Eq.~(\ref{G30}) at $q^6=q_\parallel^2 R L^{-4}$
for positive $R$ is now exactly compensated by a similar divergence
in the second term of Eq.~(\ref{G30}) for the same wave number, since
$\lambda_0=R^{1/3}$ for positive $R$. Thus
$\tilde{S}_\mathrm{NE}(\tilde{q}_\parallel,\tilde{q}_\perp)$, as given
by Eq.~(\ref{G30}), is continuous at $q^6=q_\parallel^2 R L^{-4}$
for positive $R$. However, due to the presence of a sine term in the
denominator, new divergences appear when:
\begin{equation}
\tilde{q}_\parallel^{1/3} \sqrt{R^{1/3}- \tilde{q}_\parallel^{4/3}} = N\pi
\label{G31}
\end{equation}
for $N=0,1,\dots$. For $N=0$ this new divergence is again compensated due
to the presence of the term $\sqrt{R^{1/3}-\tilde{q}_\parallel^{4/3}}$
in the numerator. But for $N=1$, there is no compensation
and the contribution~(\ref{G30}) to the nonequilibrium
enhancement of the structure factor diverges. Since
condition~(\ref{G31}) for $N=1$ is satisfied when
$R=(\tilde{q}_\parallel^2+\pi^2)^3/\tilde{q}_\parallel^2$,
this divergence will appear for any Rayleigh number
larger than the critical Rayleigh number $R_\mathrm{c}=27 \pi^4/4$.
For $R=R_\mathrm{c}$, condition~(\ref{G31}) is satisfied at a critical
wave number $\tilde{q}_{\parallel \mathrm{c}} = \pi /\sqrt{2}$.
We thus recover from an analysis of the divergences in the
nonequilibrium structure factor the well-known results of the linear
stability theory for a fluid layer bounded by two free
surfaces~\cite{Chandra}. A more detailed analysis of the behavior of our
solution for the nonequilibrium structure factor for positive $R$
near $R_\mathrm{c}$ will be presented in Section~\ref{sec:INS}.

As mentioned earlier, for the interpretation of low-angle scattering
or shadowgraph experiments one may consider the approximation
$q_\parallel \approx q$, $q_\perp \approx 0$. In the limit
$q_\perp\to 0$, Eq.~(\ref{G30}) for
$\tilde{S}_\mathrm{NE}(\tilde{q}_\parallel,\tilde{q}_\perp)$ can be
slightly simplified; in this approximation the nonequilibrium
enhancement $\tilde{S}_\mathrm{NE}(\tilde{q})$
will depend only on $\tilde{q}$, which is the experimentally
relevant quantity. In the remainder of this paper the small-angle
approximation will always be assumed. We note that
all the divergent features discussed above remain the same
in this approximation.

It is interesting to look at the asymptotic behavior of the
nonequilibrium enhancement for $q \to 0$ and for $q \to \infty$.
From Eq.~(\ref{G30}) it can be readily shown that:
\begin{equation}
\tilde{S}_\mathrm{NE} (\tilde q)\xrightarrow{\tilde{q}\to 0}~
{17\over 20160}~\tilde{q}^2,
\label{tes5:7}
\end{equation}
and
\begin{equation}
\tilde{S}_\mathrm{NE} (\tilde q)\xrightarrow{\tilde{q}\to \infty}~
{1\over\tilde{q}^4}.
\label{tes5:8}
\end{equation}
We note from Eqs.~(\ref{tes5:7}) and~(\ref{tes5:8})
that the limiting behavior of $\tilde{S}_\mathrm{NE}(\tilde{q})$
for $q \to 0$ and $q \to \infty$ does not depend on the value of the
Rayleigh number $R$.

\subsection{Interpretation of experimental results}
In accordance with Eq.~(\ref{G29}), the product
$\tilde{S}^0_\mathrm{NE}~\tilde{S}_\mathrm{NE}(\tilde{q})$
represents the nonequilibrium enhancement of the structure factor with
$\tilde{S}^0_\mathrm{NE}$ given by Eq.~(\ref{strength}) and
$\tilde{S}_\mathrm{NE}(\tilde{q})$ given by Eq.~(\ref{G30})
with $\tilde{q}_\perp \to 0$. From Eq.~(\ref{strength})
we see that the prefactor $\tilde{S}^0_\mathrm{NE}$ depends on
various thermophysical properties of the fluid and on the magnitude
of the applied temperature gradient, but $\tilde{S}^0_\mathrm{NE}$
is independent of the wave number $\tilde{q}$. Hence, to evaluate
the wave-number dependence of the nonequilibrium enhancement of the
structure factor, we only need to focus our attention on
$\tilde{S}_\mathrm{NE}(\tilde{q})$.

\begin{figure}
\begin{center}
\resizebox{0.80\textwidth}{!}{\includegraphics{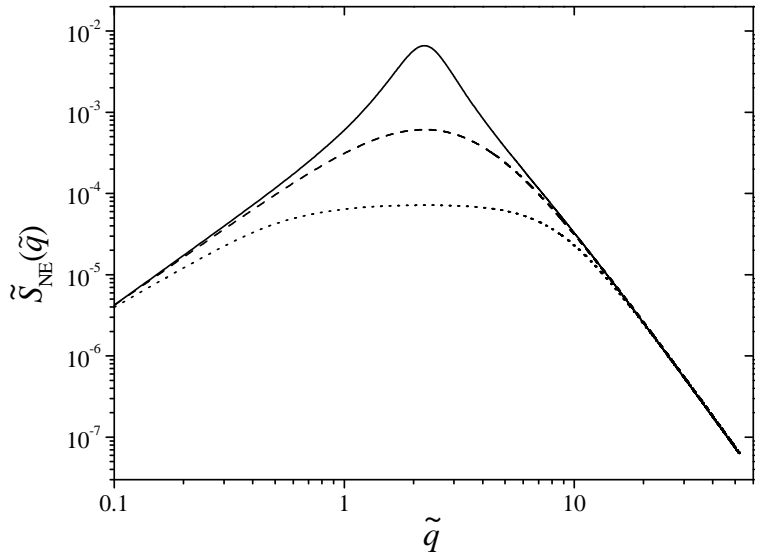}}
\end{center}
\caption{Normalized nonequilibrium enhancement $\tilde{S}_\mathrm{NE}(\tilde{q})$
of the structure factor as given by Eq.~(\ref{G30}), as a
function of the dimensionless wave number $\tilde{q}$ for three different
values of the Rayleigh number $R$. The solid curve is for $R=600$, which is
near the convective instability, the dashed curve is for $R\approx 0$ and
the dotted curve for $R=-5000$.}
\label{tercera}
\end{figure}

Assuming $\tilde{q}_\perp=0$, we have evaluated $\tilde{S}_\mathrm{NE}(\tilde{q})$
from Eq.~(\ref{G30}). In Fig.~\ref{tercera} we show on a
double-logarithmic scale  $\tilde{S}_\mathrm{NE}(\tilde{q})$ as a function of
$\tilde{q}$ for three different values of the Rayleigh number, namely for
a large negative value, $R=-5000$, for a value $R\simeq 0$ which
corresponds to the case investigated in a previous publication~\cite{Physica}
and for a value $R=600$ which is close to the critical value
$R_\mathrm{c}\simeq 656$. At larger values of $\tilde{q}$,
$\tilde{S}_\mathrm{NE}(\tilde{q})$ varies as $\tilde{q}^{-4}$, independent of the
Rayleigh number, in agreement with Eq.~(\ref{tes5:8}). Upon 
decrease of $\tilde{q}$, $\tilde{S}_\mathrm{NE}(\tilde{q})$ goes through a maximum
and for very small values of $\tilde{q}$, $\tilde{S}_\mathrm{NE}(\tilde{q})$ decreases
as $\tilde{q}^2$, in agreement with Eq.~(\ref{tes5:7}). For positive $R$,
$\tilde{S}_\mathrm{NE}(\tilde{q})$ develops a prominent peak close to
$\tilde{q}_\mathrm{c}=\pi/\sqrt{2}$, which diverges as $R\to R_\mathrm{c}$,
as further discussed in Section~\ref{sec:INS1}.
Hence, a major effect of the additive noise terms in the fluctuating
Boussinesq equations is the appearance of (fluctuating) patterns
in the fluid, even below the convective instability as discussed
by some other investigators~\cite{GarciaOjalvoBook,Staliunas}.

Sengers and coworkers have measured the nonequilibrium fluctuations
in liquid toluene~\cite{LawGammonSengers,LawEtAl,SegreEtAl1} and in
liquid $n$-hexane~\cite{Mixtures3}. These experiments correspond to
Rayleigh numbers from $-25,000$ to $-300,000$ at dimensionless wave
numbers ranging from $\tilde{q}=640$ down to $\tilde{q}=345$.
The experiments have provided an accurate confirmation of the
$q^{-4}$ dependence of the intensity of nonequilibrium fluctuations
in this range of wave numbers.

Giglio and coworkers have measured the intensity of nonequilibrium
fluctuations for negative Rayleigh numbers down to wave numbers
$\tilde{q}$ of order unity, both
with ultra-low angle light-scattering experiments~\cite{VailatiGiglio1,VailatiGiglio2}
and from quantitative analysis of shadowgraph images~\cite{GiglioNature,BrogioliEtAl}.
They actually measured the intensity of nonequilibrium {\em concentration}
fluctuations in a liquid mixture. However, due to the similar
structure of the underlying hydrodynamic equations, the $q$ dependence
of the contribution of nonequilibrium concentration fluctuations to the
structure factor in a liquid mixture is expected to be similar to the
$q$ dependence of the contribution of nonequilibrium temperature
fluctuations to the structure factor of a one-component
fluid~\cite{LectureNotes,Mexico}. Giglio and coworkers have not only confirmed the
$q^{-4}$ dependence of the nonequilibrium structure factor, but they
have also observed the crossover to a region of $\tilde{q}$ close to unity
where the nonequilibrium structure factor is independent of $\tilde{q}$,
in agreement with the flat region indicated in Fig.~\ref{tercera} for
$\tilde{S}_\mathrm{NE}(\tilde{q})$ at large negative Rayleigh
numbers.
\begin{figure}
\begin{center}
\resizebox{0.80\textwidth}{!}{\includegraphics{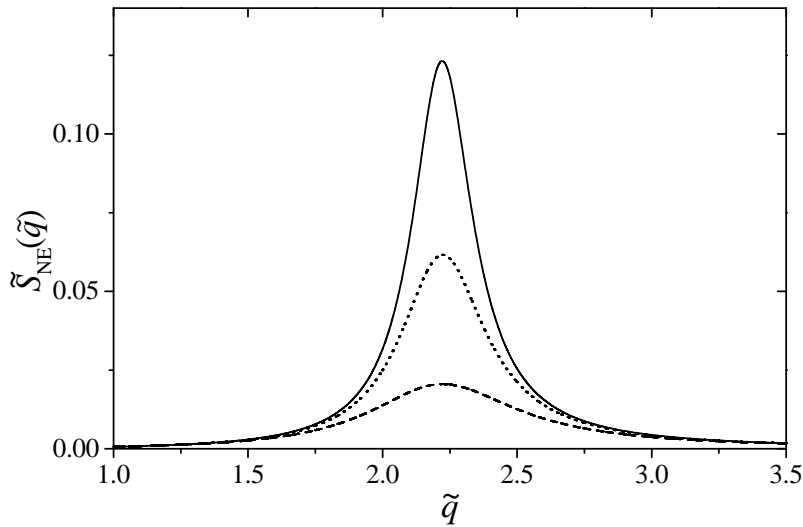}}
\end{center}
\caption{Normalized nonequilibrium enhancement $\tilde{S}_\mathrm{NE}(\tilde{q})$
of the structure factor as given by Eq.~(\ref{G30}), as a
function of the dimensionless wave number $\tilde{q}$ for three different
values of the distance $\epsilon$ to the convective instability. The solid
curve is for $\epsilon=-5\times 10^{-3}$, the dotted curve is for
$\epsilon =-10^{-2}$ and the dashed curve for $\epsilon=-3\times 10^{-2}$.}
\label{tercerab}
\end{figure}

Wu et al. have measured the nonequilibrium structure factor in a layer
of fluid carbon dioxide, at a pressure of about 3 MPa, near the convective
instability. For positive Rayleigh numbers, $\tilde{S}_\mathrm{NE}(\tilde{q})$
strongly depends on the parameter $\epsilon=(R-R_\mathrm{c})/R_\mathrm{c}$
which measures the distance from the Rayleigh-B\'{e}nard instability.
In Fig.~\ref{tercerab} we show $\tilde{S}_\mathrm{NE}(\tilde{q})$ as a function
of $\tilde{q}$ close to $\tilde{q}_\mathrm{c}=\pi/\sqrt{2}$ as calculated
from Eq.~(\ref{G30}) for $\epsilon=-0.03$, $\epsilon=-0.01$ and
$\epsilon=-0.005$. We did not attempt to make a quantitative comparison
with the measurements of Wu et al., because the slip-free boundary
conditions adopted in this paper do not correspond to the experimental
boundary conditions. However, a comparison of Fig.~\ref{tercerab} with
Fig.~3 in Ref.~\cite{WuEtAl} shows that our solution yields a good
qualitative representation of the $q$ dependence of the nonequilibrium
structure factor as measured by Wu et al. We conclude that our
solution of the linearized fluctuating Boussinesq equations for the
nonequilibrium structure factor is consistent with the characteristic
features of the nonequilibrium structure factor observed in
experiments for both negative and positive Rayleigh numbers.

\section{Nonequilibrium fluctuations close to the convective instability}
\label{sec:INS}
The nature of thermal noise near the convective instability has been
the subject of studies by many 
investigators~\cite{ZaitsevShliomis,SwiftHohenberg,KirkpatrickCohen,SchmitzCohen2,VanBeijerenCohen,HohenbergSwift}.
Hence, it is of interest to make a comparison of those results with
our solution for the intensity of temperature fluctuations
for thermal nonequilibrium states. It turns out that for an analysis
of  $\tilde{S}_\mathrm{NE}(\tilde{q})$ near the convective instability
it is more convenient not to perform the summation of the series in
Eq.~(\ref{G24}). Instead, substituting Eq.~(\ref{G24}) into
Eq.~(\ref{P31}) and integrating the terms in the series individually,
we can write $\tilde{S}_\mathrm{NE}(\tilde{q})$ as:
\begin{equation}
\tilde{S}_\mathrm{NE}(\tilde{q})= 4~
\sum_{N=1}^{\infty}
\frac{\tilde{q}^2}
{(\tilde{q}^2 + N^2 \pi^2)^3-R~\tilde{q}^2}
\frac{1-\cos(N\pi)}
{N^2\pi^2},
\label{G30b}
\end{equation}
which is equivalent to the expression~(\ref{G30}).

\subsection{Divergence of nonequilibrium fluctuations at the
convective instability\label{sec:INS1}}
Zaitsev and Shliomis~\cite{ZaitsevShliomis}
were the first to compute thermal fluctuations in a fluid
layer subjected to a stationary temperature gradient near the
convective instability. Using linear perturbation theory they found
that the structure factor diverges as $(R_\mathrm{c}-R)^{-1}$.
The same divergence follows from our solution for
$\tilde{S}_\mathrm{NE}(\tilde{q})$. To reproduce the divergence
we first calculate the value $\tilde{q}_\mathrm{max}$ corresponding
to the maximum of $\tilde{S}_\mathrm{NE}(\tilde{q})$ by expanding
$\tilde{q}_\mathrm{max}$ in powers of $\epsilon=(R-R_\mathrm{c})/R_\mathrm{c}$
around $\pi/\sqrt{2}$. Differentiating the expression~(\ref{G30b})
for $\tilde{S}_\mathrm{NE}(\tilde{q})$ and demanding the derivative to
vanish, we find that the position of the maximum corresponds to:
\begin{eqnarray}
\tilde{q}_\mathrm{max}&=&\frac{\pi}{\sqrt{2}}+
\epsilon^2~
\frac{81\sqrt{2}\pi}{8} 
\sum_{N=2}^{\infty}
\frac{(N^2-1)(1+2N^2)^2(1-\cos{N\pi})}{N^2 \left[(1+2N^2)^3-27\right]^2}
+O(\epsilon^3) \nonumber \\
&\approx&\frac{\pi}{\sqrt{2}}+6.324\times 10^{-4}~\epsilon^2.
\label{div1}
\end{eqnarray}
Substituting Eq.~(\ref{div1}) into Eq.~(\ref{G30b}) we conclude that
the structure factor, which is proportional to the intensity of the
scattered light, diverges when the convective instability is approached
such that:
\begin{equation}
\tilde{S}_\mathrm{NE}(\tilde{q}_\mathrm{max}) =
\frac{-54 \pi^2}{\epsilon}+\frac{16}{\pi^6}\sum_{N=2}^{\infty}
\frac{1-\cos{N\pi}}{N^2\left[(1+2N^2)^3-27\right]}+O(\epsilon).
\label{div2}
\end{equation}
We thus recover the linear divergence of $\tilde{S}_\mathrm{NE}$
as a function of $(R-R_\mathrm{c})^{-1}$ obtained by
Zaitsev and Shilomis~\cite{ZaitsevShliomis} and confirmed
by Swift and Hohenberg~\cite{SwiftHohenberg,HohenbergSwift}.
Linear fluctuation theory amounts to a mean-field theory of
fluctuations. Extremely close to the instability nonlinear
effects will cause a smearing out of the sharp mean-field
transition~\cite{Graham,SwiftHohenberg,HohenbergSwift}, but
this effect will only be noticeable for very small values of
$|\epsilon| \overset{<}{\sim} 2.9\times 10^{-5}$~\cite{NormandPomeauVelarde}.
Hence, observation of this linear divergence is possible in
experiments~\cite{WuEtAl}. Deviations from linear fluctuation
theory have been observed by Scherer et al. in the case
of electroconvection~\cite{SchererEtAl}.

\subsection{The most unstable mode and the Swift-Hohenberg
approximation}
The approximation scheme used by Zaitsev and Shliomis
and by Swift and Hohenberg is equivalent to retaining only the term
$N=1$ in the series expansion~(\ref{G30b}) for
$\tilde{S}_\mathrm{NE}(\tilde{q})$.
Note that, for $N=1$, when $R$ is close to $R_\mathrm{c}$ and
$q_\parallel$ is close to $q_{\mathrm{c}}$,
the denominator in the corresponding term in Eq.~(\ref{G24})
approaches zero. Therefore, in this situation the term for $N=1$
is much larger than the terms for any other value of $N$. Consequently, when
$R \overset{<}{\sim} R_\mathrm{c}$ and $q \simeq  q_{\mathrm{c}}$,
truncating the series~(\ref{G30b}) at $N=1$ yields a very good approximation.
We thus deduce from our solution for the structure factor
as measured in low-angle light scattering or shadowgraph
experiments:
\begin{equation}
S(\tilde{q}) =
\frac{8 S_\mathrm{E} \tilde{S}^0_\mathrm{NE}}{\pi^2}~
\frac{1}{\dfrac{(\tilde{q}^2 + \pi^2)^3}{\tilde{q}^2}-R},
\label{G33b}
\end{equation}
where the equilibrium contribution to $S(\tilde{q})$ has been neglected
( cf., Eq.~(\ref{G29})).
This is usually called the most-unstable-mode
approximation in the literature~\cite{ZaitsevShliomis,SwiftHohenberg}.
The denominator in Eq.~(\ref{G33b}) is zero at $R=R_\mathrm{c}=27 \pi^4/4$ and
$\tilde{q}_\parallel=\tilde{q}_{\parallel\mathrm{c}} = \pi /\sqrt{2}$.
Expanding the denominator in powers of $\tilde{q}_\parallel^{2}$ around
$\tilde{q}_{\parallel\mathrm{c}}^2$, one obtains:
\begin{equation}
S(\tilde{q}) =
\frac{4~S_\mathrm{E} \tilde{S}^0_\mathrm{NE}}{\tilde{q}^2_\mathrm{c}~R_\mathrm{c}}
\frac{1}{\tilde{\xi}_0^4~(\tilde{q}^2 -\tilde{q}_{\mathrm{c}}^2)^2 - \epsilon},
\label{G34}
\end{equation}
where, following the notation of Hohenberg and Swift~\cite{HohenbergSwift},
we have introduced the parameter $\tilde{\xi}_0^2 = 2 /(\sqrt{3} \pi^2)$.
Equation~(\ref{G34}) is proportional to the correlation function
of the Swift-Hohenberg (SH) model {\em below} the convective instability,
when nonlinear terms are negligible~\cite{SwiftHohenberg,HohenbergSwift}.
The results of SH refer to the autocorrelation of a field
$\psi(\mathbf{r}_\parallel,t)$, which is a
generalized two-dimensional order parameter introduced by them to describe
the convective instability as a nonequilibrium phase transition. 
If we compare the term $N=1$ in Eq.~(\ref{G24}) with Eq.~(2.10) in
Ref.~\cite{HohenbergSwift}, we identify our ${S}_\mathrm{NE}(q)$
as proportional to the equal-time autocorrelation function
of the field $\psi$. Therefore the SH model can be obtained as an
approximation close to the convective instability from our exact results
for $S(q_\parallel=q,q_\perp=0)$, as given by Eqs.~(\ref{G29}) and~(\ref{G30}).
As discussed by SH, Eq.~(\ref{G34}) is valid for
both ``stress-free" and ``no-slip" boundary conditions,
but the numerical values of the constants $R_\mathrm{c}$,
$q_{\parallel\mathrm{c}}$ and $\tilde{\xi}_0^2$ do depend on the
boundary conditions (see Table I in Ref.~\cite{HohenbergSwift}).
An interesting feature of the SH model
is that is a ``Lagrangian" model, because the autocorrelation
function~(\ref{G34}) can be obtained from a simple probability functional
for the field $\psi(\mathbf{r_\parallel})$, as is discussed in detail
in the relevant literature~\cite{SwiftHohenberg,HohenbergSwift}.
\begin{figure}
\begin{center}
\resizebox{0.80\textwidth}{!}{\includegraphics{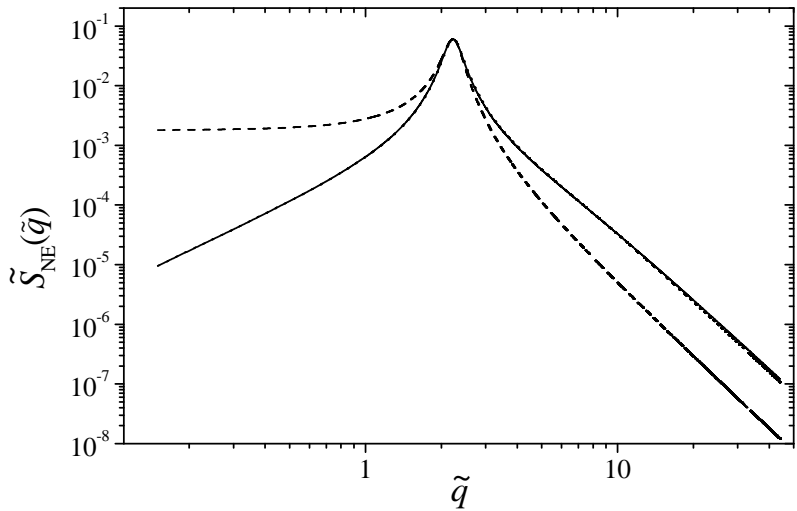}}
\end{center}
\caption{Normalized nonequilibrium enhancement $\tilde{S}_\mathrm{NE}(\tilde{q})$
of the structure factor as a function of the dimensionless
wave number $\tilde{q}$ near the convective instability, $\epsilon=-10^{-2}$.
The solid line represents the exact expression as given by Eq.~(\ref{G30}).
The dotted curve (indistingishable from the former, except asymptotically for
large $q$) represents the most-unstable-mode approximation
given by Eq.~(\ref{G33b}). The dashed curve represents the Swift-Hohenberg
approximation given by Eq.~(\ref{G34}).}
\label{cuarta}
\end{figure}

To compare our more complete solution of the fluctuating
Boussinesq equations for $\tilde{S}_\mathrm{NE}(\tilde{q})$ with
the various approximations proposed for the structure factor
close to the convective instability, we have plotted  
in Fig.~\ref{cuarta} $\tilde{S}_\mathrm{NE}(\tilde{q})$ as a function
of $\tilde{q}$. The solid line represents $\tilde{S}_\mathrm{NE}(\tilde{q})$
as given by our exact solution~(\ref{G30}), the dotted curve
(hardly visible) the most-unstable-mode approximation~(\ref{G33b})
and the dashed curve the SH-model approximation~(\ref{G34}).
All three curves correspond to a fluid layer subjected to a
temperature gradient such that $\epsilon = -0.01$. From
Fig.~\ref{cuarta} it is evident that Eq.~(\ref{G33b})
yields a very good approximation although at large $\tilde{q}$
deviations appear. The reason is that $\tilde{S}_\mathrm{NE}(\tilde{q})$
for large $\tilde{q}$ should vary as $\tilde{q}^{-4}$, while
Eq.~(\ref{G33b}) yields $(8/\pi^2) \tilde{q}^{-4}$ for large
$\tilde{q}$.

Many authors have represented the $q$ dependence of the structure
factor near the instability in terms of a Lorentzian
profile centered at $\tilde{q}_\mathrm{c}$ with a width
proportional to $\epsilon$~\cite{KirkpatrickCohen,SchmitzCohen2,VanBeijerenCohen,HohenbergSwift},
and Wu et al. have also analyzed their experimental data
in terms of a Lorentzian profile~\cite{WuEtAl}. However, our
actual solution~(\ref{G30}) does not yield a Lorentzian profile,
although it can be obtained by expanding the denominator in
Eq.~(\ref{G33b}) not in powers of $\tilde{q}^2$ around
$\tilde{q}_\mathrm{c}^2$, but in powers of $\tilde{q}$ around
$\tilde{q}_\mathrm{c}$. The problem 
is that a Lorentzian does not lead to the proper asymptotic behavior
for either small $q$ or large $q$. It even leads to an apparent
divergence when one tries to calculate the power of the fluctuations by
integrating $S(q)$ over all two-dimensional wave vectors
$q$~\cite{WuEtAl,HohenbergSwift}. We conclude that it would
be much better to represent the experimental data in terms of
an equation like Eq.~(\ref{G33b}), which is not a Lorentzian.

\subsection{Power of thermal fluctuations}
Another feature of nonequilibrium fluctuations close to the convective
instability that has been studied theoretically and experimentally
is the so-called power of the thermal fluctuations, $\langle \delta T^2 \rangle$.
The mean-square amplitude $\langle \delta T^2 \rangle$ of the
temperature fluctuations is related to the integral of the
structure factor measured in the experiments:
\begin{equation}
\langle \delta T^2 \rangle = \frac{1}{\alpha^2\rho^2}\frac{1}{L}
\int S(q) \frac{d^2\mathbf{q}}{(2\pi)^2},
\label{G35a}
\end{equation}
which corresponds to a vertical average of the real-space
temperature autocorrelation function. Due to the symmetry of the
problem this quantity does not depend on the point $\mathbf{x}_\parallel$
in the horizontal plane at which is evaluated.
In Eq.~(\ref{G35a}) the prefactor $1/(\alpha^2\rho^2)$ accounts for
the conversion from density fluctuations to temperature
fluctuations (cf., Eqs.~(\ref{DefS}) and~(\ref{G23})) and the factor
$1/L$ results from the vertical average (compare Eq.~(\ref{P31}) with
Eq.~(A14) in Ref.~\cite{HohenbergSwift}). From Eq.~(\ref{G30b})
we note that in terms of our solution of the fluctuating
Boussinesq equations (neglecting the equilibrium contribution):
\begin{equation}
\langle \delta T^2 \rangle=
\frac{S_\mathrm{E} \tilde{S}^0_\mathrm{NE}}{\alpha^2\rho^2 L \pi^2}
\sum_{N=1}^{\infty}\int_0^{\infty}
\frac{\tilde{q}^2~2 \pi q~dq}
{(\tilde{q}^2 + N^2 \pi^2)^3-R~\tilde{q}^2}
\frac{1-\cos(N\pi)}
{N^2\pi^2}.
\label{Noise2}
\end{equation}
The integral in~(\ref{Noise2}) can be performed analytically,
but the result is long and not particularly interesting.
Asymptotically close to the convective
instability we obtain:
\begin{equation}
\langle \delta T^2 \rangle \simeq
\frac{S_\mathrm{E} \tilde{S}^0_\mathrm{NE}}{\alpha^2\rho^2 L^3}
\frac{\sqrt{3}}{R_\mathrm{c} \sqrt{-\epsilon}}.
\label{Noise3}
\end{equation}
Thus the mean-square amplitude $\langle \delta T^2 \rangle$ calculated
from our solution will diverge as $1/\sqrt{-\epsilon}$, which is
the kind of divergence observed experimentally by Ahlers and
coworkers~\cite{WuEtAl,BodenschatzEtAl}.
In Eq.~(\ref{Noise3}), $\tilde{S}^0_\mathrm{NE}$ has to be
evaluated at the critical temperature gradient. It is interesting
to note that the SH approximation gives for $\langle \delta T^2 \rangle$
the same asymptotic behavior, Eq.~(\ref{Noise3}), than the exact
solution.

\section{Correlations in real space}
\label{sec:RS}
In this section we study the real-space behavior of the
autocorrelation function of the nonequilibrium density fluctuations.
We also shall make contact
with the third approach dealing with nonequilibrium fluctuations
that was mentioned in the Introduction. To calculate the autocorrelation
function in real space we apply an inverse Fourier transform in the
horizontal plane to the structure factor as given by Eq.~(\ref{G23}).
The equilibrium contribution gives a triple delta function.
The nonequilibrium contribution gives a
nontrivial contribution, which we denote as
$G_\mathrm{NE}(r_\parallel,z,z^\prime)$ and which is equal to:
\begin{equation}
\label{R1}
G_\mathrm{NE}(r_\parallel,z,z^\prime)=\frac{S_\mathrm{E}~\tilde{S}^0_\mathrm{NE}}{2 \pi}
~\int_0^\infty q_\parallel J_0(q_\parallel r_\parallel)~
\tilde{S}_\mathrm{NE}(q_\parallel,z,z^\prime)~dq_\parallel,
\end{equation}
where $J_0(x)$ is the Bessel function of the first kind and order zero.
Note that due to the cylindrical symmetry of the problem the
correlation in real space depends only on the cylindrical
radial coordinate ($r_\parallel$). It is difficult to
study the integral~(\ref{R1}) in the general case, but
there are a couple of particular cases for which integral~(\ref{R1})
can be evaluated explicitly or where simple approximations can be
made so that the real-space behavior of the correlation
function can be studied in detail.
\begin{figure}
\begin{center}
\resizebox{0.80\textwidth}{!}{\includegraphics{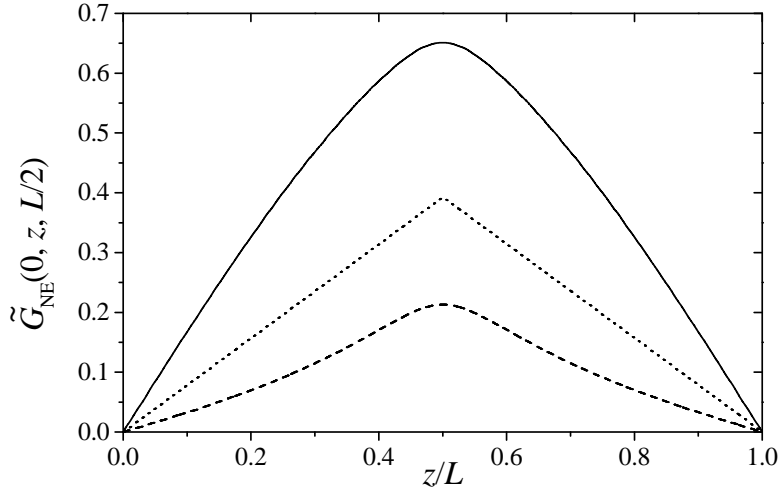}}
\end{center}
\caption{Normalized dimensionless nonequilibrium correlation function
$\tilde{G}_\mathrm{NE}(r_\parallel,z,z^\prime)$ in the Z-direction
($r_\parallel=0$) as a function of $z/L$ for $z^\prime=L/2$. The solid
curve is for $R=500$, the dotted lines are for $R=0$ and the dashed curve is
for $R=-2000$.}
\label{FigCorr1}
\end{figure}

\subsection{Correlations in the vertical direction}
The first interesting case for which the long-range
part of the real-space correlation function
$G_\mathrm{NE}(r_\parallel,z,z^\prime)$ can be studied is for
$r_\parallel=0$, i.e., along the vertical Z-direction. In this
particular case, the correlation function $G_\mathrm{NE}$ will
only depend on the variables $z$ and $z^\prime$ and on the
Rayleigh number $R$. Close to the convective instability, the
behavior of the real-space correlation function will be dominated
by the most unstable mode. From Eq.~(\ref{G24}) we note that $G_\mathrm{NE}$
will be proportional to the product $\sin(\pi z/L)~\sin(\pi z^\prime/L)$.
Therefore, the concavity of $G_\mathrm{NE}(0,z,z^\prime)$ as a
function of $z$ for fixed value of $z^\prime$ is negative. For the
interesting case $R=0$, $G_\mathrm{NE}(0,z,z^\prime)$
was evaluated exactly in a previous publication~\cite{Physica}.
It turns out that the concavity of $G_\mathrm{NE}(0,z,z^\prime)$
changes at $R=0$, being negative for positive $R$ and positive
for negative $R$ (except for a small region near $z=z^\prime$).
In Fig.~\ref{FigCorr1} we have plotted the dimensionless nonequilibrium
correlation function, $\tilde{G}_\mathrm{NE}=G_\mathrm{NE} L^3 /(S_\mathrm{E} \tilde{S}_\mathrm{NE}^0)$,
as a function of $z$, for $z^\prime=L/2$, $r_\parallel=0$ and for three
values of the Rayleigh
number $R$. The dotted lines correspond to $R=0$, the solid curve
corresponds to a positive $R$ and the dashed curve to a negative $R$.
The change of the concavity
of the function at $R=0$ is evident. This feature can be investigated
analytically since, for small values of the Rayleigh number,
$G_\mathrm{NE}(0,z,z^\prime)$ can be readily expanded in powers of
$R$ yielding:
\begin{align}
G_\mathrm{NE}(0,z,z^\prime) \simeq
\frac{S_\mathrm{E} \tilde{S}_\mathrm{NE}^0}{2\pi L^3} & \left\{
\frac{|z+z^\prime|-|z-z^\prime|}{8 L}-\frac{z z^\prime}{4 L^2} \right.
\nonumber \\ & + \left.
\frac{R}{1350} \left[ B_6\left(\frac{|z-z^\prime|}{2L}\right) -
B_6\left(\frac{|z+z^\prime|}{2L}\right) \right] \right\},
\label{RS1}
\end{align}
where $B_6(x)$ is the Bernoulli polynomial of the 6th degree (actually,
it is possible to obtain a systematic expansion of 
$G_\mathrm{NE}(0,z,z^\prime)$ in terms of the Bernoulli polynomials,
but the resulting series is not very informative). For $R=0$, we recover
from Eq.~(\ref{RS1}) the expression obtained for the spatial dependence
of the correlation function in a previous publication
(see Eq.~(25) and Fig~2 in Ref.~\cite{Physica}).
The change of concavity of $G_\mathrm{NE}(0,z,z^\prime)$ at
$R=0$ is also evident from Eq.~(\ref{RS1}). We note that for $R=0$ the
function $G_\mathrm{NE}(0,z,z^\prime)$ has a kink at $z=z^\prime$,
indicating the presence of a discontinuity in the derivative at this
point. For any other value of $R<R_\mathrm{c}$ the function and its
derivative are continuous.

From Fig~\ref{FigCorr1} we also note that the nonequilibrium
real-space equal-time correlation function is always positive, has a
maximum at $z=z^\prime$ and
decreases monotonically from this maximum reaching zero at both
ends of the $[0,L]$ interval. Thus, we conclude that the real-space
correlation function in the vertical direction is nonlocal, long ranged
and does not involve any intrinsic length scale, i.e., the correlation
encompasses the entire system only to be cut off by the finite
size of the system itself. Hence, the presence of the temperature
gradient induces long-range correlations in the direction coincident with
the temperature gradient, as has been elucidated by previous investigators,
mainly on the basis of the nonequilibrium heat-diffusion
equation~\cite{GarciaEtAl1,TornerRubi,PagonabarragaEtAl}.
Considering only the heat-diffusion equation means that one neglects any
coupling between temperature and velocity fluctuations.
This approach neglects also any gravity effects and is restricted to
fluctuations with $q_\parallel=0$.
The existence of long-range
correlations in the direction of the temperature gradient has been verified
by numerical integration~\cite{GarciaEtAl2}, by computer
simulation~\cite{MalekMansourEtAl2}, by lattice-gas-automaton
dynamics~\cite{SuarezEtAl} and by a master-equation
approach~\cite{BreuerPetruccione}. Our result for the
correlation function along the direction of the temperature gradient
is in agreement with that obtained by Malek Mansour and
coworkers~\cite{GarciaEtAl1,GarciaEtAl2}, as is also evident form a
comparison of our Fig.~\ref{FigCorr1} for $R=0$ with Fig.~1 in
Ref.~\cite{GarciaEtAl2}. Some of these authors have also considered a
coupling between temperature and velocity
fluctuations~\cite{MalekMansourEtAl1}. But to simplify the hydrodynamic
equations they made the assumption that the thermal expansion coefficient
of the fluid vanishes. This is equivalent to assuming that density
fluctuations are caused only by pressure fluctuations. Hence, their
calculations are concerned with Brillouin scattering, while here we are
considering Rayleigh scattering, for which
accurate experimental information on nonequilibrium fluctuations has been
obtained. More recently, Bena et al. have studied the
velocity autocorrelation function adopting an hydrodynamic approximation
known as Kolmogorov flow~\cite{BenaEtAl}; again, due to the nature of the
hydrodynamic simplifications performed, the results cannot be compared with
those obtained in the present paper.

\begin{figure}
\begin{center}
\resizebox{0.80\textwidth}{!}{\includegraphics{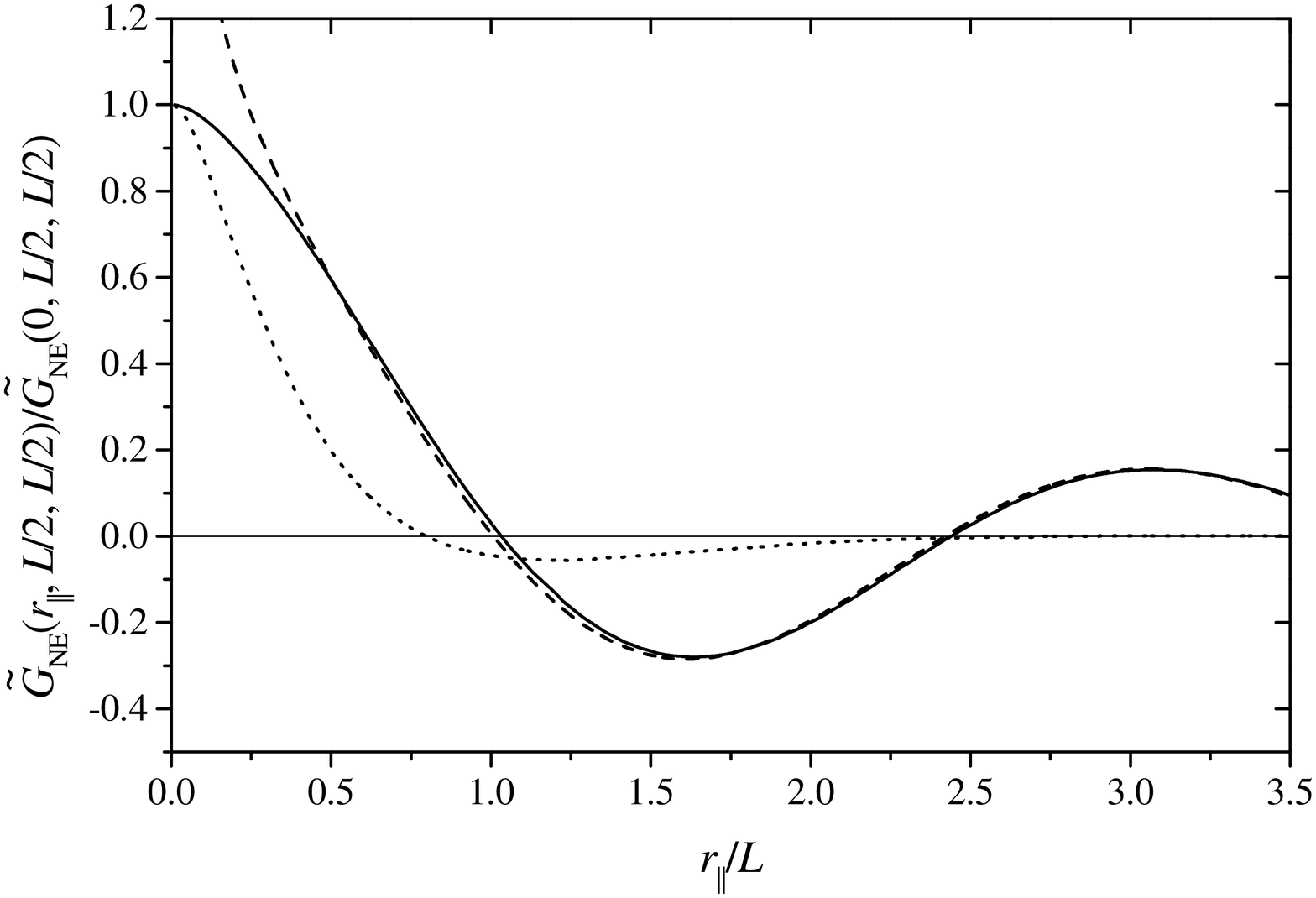}}
\end{center}
\caption{Normalized dimensionless nonequilibrium correlation function
$\tilde{G}_\mathrm{NE}(r_\parallel,z,z^\prime)$ in the central plane
of the liquid layer ($z=z^\prime=L/2$), near the convective instability,
as a function of $r_\parallel/L$, divided by its value at $r_\parallel=0$.
The solid curve is for $\epsilon=-10^{-2}$; the dashed curve is the asymptotic
approximation, Eq.~(\ref{R6}), for the same value of $\epsilon$. The dotted
curve is for $\epsilon=-0.7$.}
\label{FigCorr2}
\end{figure}
\subsection{Correlations in the horizontal plane}
A second particular case for which the real-space behavior of the
correlations in a liquid subjected to a stationary temperature gradient
can be
easily studied is in a horizontal plane, that is, in the plane perpendicular
to the temperature gradient and far from the boundaries ($z=z^\prime=L/2$).
We shall analyze here the spatial dependence of the correlations
in the horizontal direction near the convective instability using the
most-unstable-mode approximation, Eq.~(\ref{G33b}),
which, as was shown in Section~\ref{sec:INS}, yields an excellent
representation of the structure factor near the convective instability.
In this approximation:
\begin{equation}
\label{R5}
G_\mathrm{NE}(r_\parallel,L/2,L/2)=\frac{S_\mathrm{E}~\tilde{S}^0_\mathrm{NE}}{2 \pi^2 L^3}
\int_0^\infty
\frac{\tilde{q}_\parallel^3 J_0(\tilde{q}_\parallel \tilde{r}_\parallel)}
{(\tilde{q}_\parallel^2+\pi^2)^3-R \tilde{q}_\parallel^2}
~d\tilde{q}_\parallel,
\end{equation}
where $\tilde{r}_\parallel = r/L$.
The integral (\ref{R5}) can be evaluated exactly, when $R<R_\mathrm{c}$, but the
result is a rather long expression. Instead we prefer to plot in
Fig.~\ref{FigCorr2} the resulting
$G_\mathrm{NE}(r_\parallel,z=z^\prime=L/2)/G_\mathrm{NE}(0,z=z^\prime=L/2)$
as a function of $r_\parallel$ for two values of the Rayleigh
number. The solid curve represents the density
autocorrelation function for $\epsilon=-0.01$, which is close to the
convective instability; the dotted curve represents the same
function, but for $\epsilon=-0.7$. For $R$ well below $R_\mathrm{c}$, the
correlation function is overdamped but, as the value of $R$ approaches
$R_\mathrm{c}$, it starts to oscillate and the number of oscillations 
before the correlation function decays to zero increases the closer
one approaches $R_\mathrm{c}$.
To see this mathematically, we have made an asymptotic expansion
of $G_\mathrm{NE}(r_\parallel,z=z^\prime=L/2)$ for large $r_\parallel$
and, when $R$ is close to $R_{c}$, we obtain:
\begin{eqnarray}
G_\mathrm{NE}(r_\parallel,L/2,L/2)&\simeq&
\frac{S_\mathrm{E}~\tilde{S}^0_\mathrm{NE}}{81 2^{1/4} \pi^3 L^3}
\sqrt{\frac{108-151\epsilon}{-\epsilon \tilde{r}_\parallel}}
\exp{\left(\frac{-\sqrt{-6\epsilon}}{4} \pi \tilde{r}_\parallel \right)} \nonumber \\
&\times&\cos{\left[\frac{\sqrt{2}\pi \tilde{r}_\parallel}{2}-\left(
\frac{\pi}{4} - \frac{23\sqrt{-3\epsilon}}{36}\right)\right]}.
\label{R6}
\end{eqnarray}
The approximation~(\ref{R6}) for $\epsilon=-0.01$ is shown in
Fig.~\ref{FigCorr2} as a dashed curve, to be compared with the solid
curve representing the autocorrelation function as given by
Eq.~(\ref{R5}) for the same value of $\epsilon$.
As can be inferred from
Fig.~\ref{FigCorr2}, the approximation is excellent for small $\epsilon$
and for large $\tilde{r}_\parallel$. From Eq.~(\ref{R6}), the
real-space behavior of the autocorrelation function in the horizontal plane
as a kind of damped oscillation is evident.

\section{Summary and concluding remarks}
\label{sec:CON}
Starting from the linearized fluctuating Boussinesq equations we have
derived the structure factor of a horizontal fluid layer taking into
account both gravity and finite-size effcts. The resulting expression
reproduces the $q^{-4}$ dependence of the nonequilibrium structure factor
predicted theoretically~\cite{KirkpatrickEtAl,RonisProcaccia,SchmitzCohen1}
and confirmed experimentally~\cite{LawGammonSengers,LawEtAl,SegreEtAl1,LectureNotes}
for negative Rayleigh numbers, it accounts for the saturation of the
nonequilibrium enhancement of the intensity of the fluctuations
at small wave numbers observed by Giglio and coworkers~\cite{VailatiGiglio1,VailatiGiglio2,GiglioNature,BrogioliEtAl}
and it is consistent with the experimental observation by
Ahlers and coworkers~\cite{WuEtAl,BodenschatzEtAl} of the structure
factor close to the convective instability. We have thus provided a
unified approach for describing nonequilibrium fluctuations for both
negative and positive values of the Rayleigh number, provided that
$R<R_\mathrm{c}$. Our solution also enables us to evaluate the real-space
behavior of the correlation function of the temperature fluctuations
which agrees with the results obtained by other investigators for
the spatial direction co-incident with the direction of
gravity~\cite{GarciaEtAl1,GarciaEtAl2,SuarezEtAl,BreuerPetruccione,MalekMansourEtAl1}.

There are still a number of issues that deserve further attention.

First of all, for the sake of mathematical simplicity, we have adopted
stress-free boundary conditions in the present paper. For a quantitative
comparison with experimental data at very small wave numbers and close to
the convective instability modifications due to no-slip boundary
conditions should be considered~\cite{EPJ,SchmitzCohen2,AhlersEtAl,VanBeijerenCohen,HohenbergSwift}.

Second, we want to emphasize that neither our solution nor that of
Swift and Hohenberg (SH) close to the convective instability yield a
Lorentzian wave number dependence on $q$ as is often assumed in the
literature since a Lorentzian does not yield the proper
asymptotic behavior for either small or large $q$.
Moreover, even when the SH equation is not approximated by a Lorentzian,
it does not yield the expected asymptotic $q$ behavior either.
We plan to study this
issue in more detail in the future.

Finally, while the average local values $\langle\theta\rangle$
and $\langle w\rangle$ are zero, the average cross correlation
$\langle\theta\cdot w\rangle$ outside equilibrium is not generally 
zero. As a result, the coupling between temperature and transverse-velocity
fluctuations may cause a small convective heat transfer, even below the
instability~\cite{AhlersEtAl,VanBeijerenCohen}. A satisfactory solution
of this effect has not yet been obtained~\cite{MeyerEtAl1,VanBeijerenCohen,AhlersEtAl2,MeyerEtAl2}.

\section*{Acknowledgments}
The authors acknowledge some valuable discussions with
G. Ahlers, D.S. Cannell, T.R. Kirkpatrick, R. P\'{e}rez Cord\'{o}n
and M.G. Velarde. The research at the University of Maryland is supported
by the Chemical Sciences, Geosciences and Biosciences Division of the Office
of Basic Energy Sciences of the US Department of Energy under Grant
No. DE-FG-02-95ER14509. J.V.S. acknowledges the hospitality of the
Departamento de F\'{\i}sica Aplicada I, Universidad Complutense de Madrid,
where the manuscript was completed.

\end{document}